\documentclass[11pt]{article}
\pdfoutput=1 

\usepackage{jheppub} 

\usepackage[utf8]{inputenc}
\usepackage[english]{babel}
\usepackage{amsmath,amssymb,amsbsy,amstext,amsthm,booktabs,simplewick,exscale,re
lsize,slashed,graphicx,amsfonts,upgreek, xcolor}
\usepackage{tabu,multirow,color,dcolumn,bm,enumerate}
\usepackage{hyperref}
\usepackage{feynmp-auto,axodraw2,tikz}

\newcommand{\eq}[1]{\begin{equation} #1 \end{equation}}
\newcommand{\eql}[2]{\begin{equation} \label{eqn:#1} #2 \end{equation}}
\newcommand{\eqs}[1]{\begin{equation} \begin{split} #1 \end{split} \end{equation}}
\newcommand{\eqsl}[2]{\begin{equation} \label{eqn:#1} \begin{split} #2 \end{split} \end{equation}}

\newcommand{\nl}{\nonumber \\}
\newcommand{\ra}{\rightarrow}

\newcommand{\eqnref}[1]{Eq.~(\ref{eqn:#1})}

\newcommand{\secref}[1]{Sec.~\ref{sec:#1}}

\newcommand{\appref}[1]{Appendix~\ref{app:#1}}

\newcommand{\figref}[1]{Fig.~\ref{fig:#1}}

\newcommand{\wt}[1]{\widetilde{#1}} 
\newcommand{\oln}[1]{\overline{#1}} 

\newcommand{\ld}[1]{_{\mathrm{#1}}}
\newcommand{\mbf}[1]{\mathbf{#1}} 
\newcommand{\mcal}[1]{\mathcal{#1}} 
\newcommand{\mrm}[1]{\mathrm{#1}} 

\newcommand{\pr}[1]{\left(#1\right)}
\newcommand{\br}[1]{\left[#1\right]}
\newcommand{\crl}[1]{\left\{#1\right\}}
\newcommand{\abs}[1]{\left|#1\right|}

\newcommand{\fracp}[3]{\left( \frac{#1}{#2} \right)^{#3}}
\newcommand\Tr{\mathrm{Tr}} 

\newcommand\pd{\partial} 

\def\a{\alpha}
\def\b{\beta}

\def\d{\delta}
\def\e{\epsilon}
\def\vare{\varepsilon}

\def\varf{\varphi}
\def\g{\gamma}

\def\j{\psi}
\def\k{\kappa}
\def\l{\lambda}
\def\o{\omega}
\def\q{\theta}
\def\r{\rho}
\def\s{\sigma}
\def\t{\tau}

\def\D{\Delta}

\def\G{\Gamma}

\def\O{\Omega}

\title{Effective Field Theory of St\"uckelberg Vector Bosons}
\author[a]{Graham D. Kribs,}
\author[b,c,d]{Gabriel Lee,}
\author[e]{and Adam Martin}

\affiliation[a]{Institute for Fundamental Science and Department of Physics, \\
  University of Oregon, Eugene, OR, 97403 USA}
\affiliation[b]{Department of Physics, LEPP, Cornell University,
  Ithaca, NY, 14853, USA} 
\affiliation[c]{Department of Physics, Korea University, Seoul 136-713, Korea} 
\affiliation[d]{Department of Physics, University of Toronto,
  Toronto, ON, Canada} 
\affiliation[e]{Department of Physics, University of Notre Dame,
  South Bend, IN, 46556 USA}

\emailAdd{kribs@uoregon.edu}
\emailAdd{leeg@korea.ac.kr}
\emailAdd{amarti41@nd.edu}

\abstract{
  We explore the effective field theory of a vector field $X^\mu$
  that has a St\"uckelberg mass.  
  The absence of a gauge symmetry for $X^\mu$ implies Lorentz-invariant
  operators are constructed directly from $X^\mu$.
  Beyond the kinetic and mass terms, allowed interactions
  at the renormalizable level include 
  $X_\mu X^\mu H^\dagger H$,
  $(X_\mu X^\mu)^2$,
  and $X_\mu j^\mu$, where $j^\mu$ is a global current of the SM
  or of a hidden sector. 
  We show that all of these interactions lead to scattering amplitudes that
  grow with powers of $\sqrt{s}/m_X$, except for
  the case of $X_\mu j^\mu$ where $j^\mu$ is 
  a \emph{nonanomalous} global current. 
  The latter is well-known when $X$ is identified as a dark photon
  coupled to the electromagnetic current, often written equivalently
  as kinetic mixing between $X$ and the photon. 
  The power counting for the energy growth of the scattering
  amplitudes is facilitated
  by isolating the longitudinal enhancement.
  We examine in detail the interaction with an \emph{anomalous}
  global vector current $X_\mu j_{\rm anom}^\mu$,
  carefully isolating the finite contribution to the
  fermion triangle diagram.
  We calculate the longitudinally-enhanced 
  observables 
  $Z \rightarrow X\gamma$ (when $m_X < m_Z$), 
  $f\bar{f} \rightarrow X \gamma$,
  and $Z\gamma \to Z\gamma$
  when $X$ couples to the baryon number current.  
  Introducing a ``fake'' gauge-invariance by writing 
  $X^\mu = A^\mu - \partial^\mu \pi/m_X$, the
  would-be gauge anomaly associated with $A_\mu j_{\rm anom}^\mu$
  is canceled by $j_{\rm anom}^\mu \partial_\mu \pi /m_X$;
  this is the four-dimensional Green--Schwarz anomaly-cancellation
  mechanism at work.
  Our analysis demonstrates there is a much larger set of possible
  interactions that an EFT with a St\"uckelberg vector field can have,
  revealing scattering amplitudes that grow with energy.  
  The growth of these amplitudes can be tamed by a dark Higgs sector,
  but this requires dark Higgs boson interactions (and reintroduces
  fine-tuning in the dark Higgs sector) that can be
  separated from $X$ interactions only in the limit $g \ll 1$.
}
\begin{document}
\maketitle

\setcounter{page}{2}

\section{Introduction}
\label{sec:intro}

New massive vector bosons are ubiquitous in beyond the
Standard Model (SM) physics.  At masses large compared with
collider energies, they provide UV completions of higher
dimensional operators \cite{Brivio:2017vri}.
At intermediate masses, of order collider energies, they yield
resonances that are targeted by many searches \cite{Langacker:2008yv}.
At somewhat smaller masses, they can be produced, decay, 
and be observed in high intensity experiments 
\cite{Batell:2009di,Bjorken:2009mm,Jaeckel:2010ni},
typically when coupled to charged leptons
(for reviews, see \cite{Alexander:2016aln,Battaglieri:2017aum}).  
Also at smaller masses, they can act as mediators to permit
light dark matter to interact with the SM
\cite{Arkani-Hamed:2008hhe,Pospelov:2008jd,Pospelov:2008zw},
underpinning the viability of a large class of light dark matter 
detection experiments \cite{Battaglieri:2017aum}. 
At exceptionally small masses, vector bosons
can even serve as dark matter itself
\cite{Nelson:2011sf,Arias:2012az,Graham:2015rva,Agrawal:2018vin,Co:2018lka,Bastero-Gil:2018uel,Dror:2018pdh,Co:2021rhi}.

One of the attractions of a single new massive vector boson is
that a simple model \cite{Holdom:1985ag} exists: 
the massive $U(1)$ dark photon $A^\mu$ 
(see \cite{Fabbrichesi:2020wbt} for a review),
\eql{darkphoton}{
  \mathcal{L}_{\text{dark } \g} =
  - \frac{1}{4} F_{A,\mu\nu} F_A^{\mu\nu}
  + \frac{1}{2} m_X^2 A_{\mu} A^\mu                                 
  - \epsilon F_{A,\mu\nu} F_Y^{\mu\nu} \,,
}
that involves just two parameters $m_X$ and $\epsilon$, 
respectively the mass of the $U(1)$ dark photon 
and its kinetic mixing to hypercharge. 
The simplicity of this extension hinges on the existence of a St\"uckelberg mass
(see \cite{Ruegg:2003ps} for a review) for the dark photon.
In particular, by not specifying a Higgs mechanism for the dark photon, 
one is able to avoid the consideration of additional interactions of the
dark Higgs field $\phi_X$.
In particular, one does not need to address the new fine tunings from the 
``dark hierarchy problem'' that are inevitable with a dark Higgs field
or how to avoid the respective destabilization of the dark and/or
SM Higgs sectors through renormalizable interactions such as
$\phi_X^\dagger \phi_X H^\dagger H$.

One of the reasons the dark photon Lagrangian \emph{seems} simple
is how the longitudinal mode is packaged in $A^\mu$.
We can introduce the longitudinal mode $\pi$ such that,
under a gauge transformation
$A^\mu \ra A^\mu + \partial^\mu \a(x)$,
the longitudinal mode shifts 
$\pi \ra \pi + m_X \a(x)$.
Using the equation of motion (EOM) for the hypercharge gauge boson,
$\partial_\mu F^{\mu\nu}_Y = g_Y j^\nu_Y$
in terms of the SM hypercharge current $j^\nu_Y$, 
the dark photon Lagrangian can be rewritten as:
\eql{darkphotonX}{
  \mathcal{L}_{\text{dark } \g} =
  - \frac{1}{4} F_{X,\mu\nu} F_X^{\mu\nu}
  + \frac{1}{2} m_X^2 X_\mu X^\mu                                 
  - \epsilon g_Y X_\mu j^\mu_Y
}
in terms of $X^\mu \equiv A_X^\mu - \partial^\mu \pi/m_X$ -- 
the St\"uckelberg vector field -- a vector boson without a
corresponding $U(1)$ gauge invariance.  The lack of gauge invariance
is obvious because $X^\mu$ remains invariant under the simultaneous
gauge transformations of $A^\mu$ and $\pi$.  This form of the dark photon
Lagrangian makes it clear that a Lagrangian with a
St\"uckelberg mass for a vector field is best expressed in terms of
$X^\mu$; the use of the field strength $F_X^{\mu\nu}$ for the kinetic term 
(or kinetic mixing with the SM) has nothing to do with gauge invariance, 
and instead simply ensures there are only three
propagating degrees of freedom (DOF) in $X^\mu$.%
\footnote{Contrast this with a spin-one gauge field, such as hypercharge $B^\mu$,
  which only appears in the field strength $F_B^{\mu\nu}$ and
  covariant derivatives.}

This naturally leads to the question of the effective field theory
involving a St\"uckelberg vector field $X^\mu$ --- what are
all possible interactions of $X^\mu$, and what are their consequences?
The goal of this paper is to show that the Lagrangian
\eqnref{darkphotonX} is a special case of a more general
set of interactions for $X^\mu$.  For instance, already at the
renormalizable level we can write $(X_\mu X^\mu)^2$,
$X_\mu X^\mu H^\dagger H$, and $X_\mu j^\mu$ where $j^\mu$
is a global vector or axial current that may or may not be (globally)
anomaly-free.%
\footnote{The phenomenological implications of the quartic interaction 
  for the electromagnetic field was explored in \cite{Reece:2018zvv}.}
As we will see, most of these interactions have couplings 
of the longitudinal mode with itself or the SM fields, and thus lead
to scattering amplitudes that grow with powers $\sqrt{s}/m_X$.
This is analogous to the energy growth that arise in a
Higgsless SM \cite{Lee:1977eg}. 
The range of validity of the effective theory including $X$ in the spectrum 
relies on taking the coefficients of longitudinally-enhanced interactions to be
(sometimes exceptionally) small.
Only if there are \emph{exactly zero} couplings 
of the longitudinal mode with itself or with the SM can
the cutoff scale of the EFT be taken arbitrarily large relative to the
mass of the St\"uckelberg vector field.

There is a host of related literature that we will only briefly mention.  
Numerous papers have studied theories with a St\"uckelberg vector field
in the context of field theory or string theory
\cite{Delbourgo:1987np,%
Bodwin:1991xt,%
Grosse-Knetter:1992tbp,%
Ruegg:2003ps,%
Kors:2004dx,%
Kors:2005uz,%
Coriano:2007fw,%
Burgess:2008ri,%
Goodsell:2009xc,%
Heisenberg:2014rta,%
BeltranJimenez:2016rff,%
Reece:2018zvv,%
Kachanovich:2021eqa}.
There is also a huge literature on anomalous $U(1)$ symmetries and
their implications for theory or phenomenology
\cite{Preskill:1990fr,%
Anastasopoulos:2006cz,%
Coriano:2007fw,%
Harvey:2007ca,%
Coriano:2008pg,%
Armillis:2008bg,%
Racioppi:2009yxa,%
Antoniadis:2009ze,%
Dobrescu:2014fca,%
Dror:2017ehi,%
Ismail:2017ulg,%
Dror:2017nsg,%
Ekstedt:2017tbo,%
Craig:2019zkf,%
Dror:2020fbh,%
Allanach:2020zna,%
Michaels:2020fzj,%
Bonnefoy:2020gyh,%
Ekhterachian:2021rkx,%
Chen:2021qaf}.
The connections between anomalous $U(1)$ symmetries and the Green--Schwarz 
anomaly cancellation mechanism have also been elucidated
\cite{Anastasopoulos:2006cz,%
Kumar:2007zza,%
Coriano:2007fw,%
Coriano:2008pg,%
Armillis:2008bg,%
Ekstedt:2017tbo}.  
While we have certainly benefited from this literature and we do not claim 
to be the first or last word on this subject, our focus on 
a theory with a St\"uckelberg mass for $X^\mu$, a vector field
without a corresponding gauge symmetry, lays a foundation for 
a systematic approach to analyze the effective field theory
of $X^\mu$ in terms of its leading self-interactions as well as its
interactions with the SM\@.

The organization of this paper is as follows.  
First, in \secref{reviewmassivevector}, we review the St\"uckelberg Lagrangian,
(fake) gauge fixing, BRST, the external physical states, the propagator,
and the BRST current.  
In \secref{treelevelstuckelberg} we consider tree-level
interactions of the St\"uckelberg vector field $X^\mu$.
We demonstrate that self-couplings as well as tree-level couplings
of the longitudinal mode with the SM lead to amplitudes that
grow with energy above the mass of the St\"uckelberg vector field.
While these interactions are not radiatively generated
by a dark photon Lagrangian that consists solely of a mass term
and a coupling to a conserved vector current, there are no symmetries
that forbid these terms. Consequently, the dark photon Lagrangian 
appears rather peculiar. In particular, we show that these interactions can be 
generated by a dark Higgs mechanism for a dark $U(1)$ gauge theory, 
and like the Higgs mechanism of the SM, the dark Higgs boson renders the 
amplitudes finite above the dark Higgs mass.  
In \secref{anomalous}, we consider the coupling of a
St\"uckelberg vector field to an \emph{anomalous} vector current.
This is motivated by Dror et al.~\cite{Dror:2017nsg},
who showed that should an anomalous symmetry of the SM 
(e.g., baryon number) be gauged,
the couplings of the longitudinal mode lead to longitudinal
enhancements of the amplitudes involving the anomalous fermion
triangle diagram. 
These longitudinal enhancements are critical in determining the viable range 
of parameter space in the model \cite{Dror:2017ehi}. 
The St\"uckelberg vector field theory would appear to be special,
since there is no gauge symmetry, and thus, no gauge anomalies.  
Nevertheless, we carefully consider the one-loop triangle diagrams 
that arise because of an anomalous \emph{global} symmetry of the SM\@.  
We find that the St\"uckelberg vector field has couplings of its
longitudinal mode to the divergence of the anomalous global current.  
The observable predictions of a St\"uckelberg vector field coupled to, 
say, global baryon number of the SM are identical to the case in which 
baryon number is gauged, so long as the ``anomalons'' needed to cancel
the gauge anomaly are taken to be heavy.
In \secref{pheno}, we demonstrate the importance of the one-loop couplings of
the longitudinal part of $X^\mu$
to an anomalous global current for several physical
processes, including $Z \ra X \gamma$ and $f \bar{f} \ra X \gamma$,
and $Z\gamma \ra Z\gamma$, when $X$ couples to baryon number.
Finally, in \secref{discussion}, we discuss the implications of our results.
The appendices contain technical details of calculations relevant for results in 
\secref{anomalous} and \secref{pheno}.

\section{Review of quantization of massive vector fields}
\label{sec:reviewmassivevector}

\subsection{The Lagrangian and propagator for a massive spin-one field}

A massive spin-one field $X^\mu$ has three propagating degrees of freedom~(DOF).
We see this by decomposing the four components of the four-vector $X^\mu$ into the 
$\mbf{1} \oplus \mbf{3}$, or spin-zero and spin-one, representations of the Lorentz group.
The spin-zero component leads to a negative energy density, 
and can be removed as a propagating DOF in the theory by imposing the Lorenz condition%
\eq{
\pd_\nu X^\nu = 0 \,,
}
together with writing the kinetic term for the four-vector as a function of the field-strength tensor 
$F_X^{\mu\nu} = \partial^\mu X^\nu - \partial^\nu X^\mu$ \cite{WeinbergVol1}.
The above two requirements are achieved by the Proca Lagrangian
\eql{LProca}{
\mathcal{L}\ld{P} = - \frac{1}{4} F_{X,\mu\nu} F_X^{\mu\nu} + \frac{1}{2} m_X^2 X_\mu X^\mu \,,
}
which yields the EOM and its derivative
\eqs{
\pd_\mu F_X^{\mu\nu} + m_X^2 X^\nu &= 0 \,, \\
m_X^2 \pd_\nu X^\nu &= 0 \,.
}
For $m_X \neq 0$, the Lorenz condition follows from the second line and therefore is not an independent constraint.
The Proca Lagrangian for $X^\mu$ is not gauge invariant: there is 
no $U(1)$ symmetry associated with $X^\mu$ since there is no redundancy
in its description---all three of its propagating DOF are physical.

The propagator for $X^\mu$ can be derived directly from inverting the
Proca Lagrangian, which is textbook material \cite{WeinbergVol1,Banks:2014twn}
\eqsl{X2pt1}{
\langle X^\mu(p) X^\nu(-p) \rangle
= \frac{-i}{p^2 - m_X^2} 
      \left( g^{\mu\nu} - \frac{p^\mu p^\nu}{m_X^2} \right) \, .
}
The propagator for $X^\mu$ is equivalent to the propagator of a Higgsed,
massive $U(1)$ theory in unitary gauge;
however, we emphasize that the result above 
is \emph{not} in unitary gauge---there is no gauge invariance.
This also implies that the sum of the polarization states for an
on-shell $X^\mu$ coincides with that of a massive $U(1)$ theory, i.e., 
\eq{
  \sum_\lambda \epsilon_\lambda^\mu(p) {\epsilon_\lambda^\nu}^*(p)
  = - \left( g^{\mu\nu} - \frac{p^\mu p^\nu}{m_X^2} \right) \, .
}
This explicitly demonstrates the counting of the on-shell physical DOF: 
$X^\mu$ has three physical polarizations.

\subsection{St\"uckelberg formalism: introducing a fake gauge symmetry} 
\label{sec:fakegauge}

The St\"uckelberg formalism expresses
\eql{XApi}{
X^\mu \equiv A^\mu - \frac{\partial^\mu \pi}{m_X} \,,
}
where $A^\mu$ is a ``fake'' $U(1)$ gauge field and $\pi$ is a scalar field
that also transforms under this ``fake'' $U(1)$ gauge invariance: 
\eqsl{Stucktransf}{
A^\mu &\to A^\mu + \partial^\mu \alpha(x) \,, \\
\pi &\to \pi + m_X \alpha(x) \,,
}
where $\alpha(x)$ is the gauge parameter.  
The Proca Lagrangian becomes
\eql{LApi}{
\mathcal{L}_g = - \frac{1}{4} F_{A,\mu\nu} F_A^{\mu\nu} 
+ \frac{1}{2} m_X^2 \left( A_\mu -\frac{\partial_\mu \pi}{m_X} \right)^2 \,,
}
purely in terms of the ``fake'' gauge field with its
its field strength given by $F_A^{\mu\nu}$.
While this construction introduces one additional DOF $\pi$,
the ``fake'' $U(1)$ gauge invariance removes one DOF,
leaving the same three of the massive vector field in the original
Proca Lagrangian \cite{Pauli:1941zz,Glauber:1953}.

We use the term ``fake'' to describe the gauge invariance of 
$A^\mu$ since the physical consequences of $X^\mu$ and its
interactions can be determined entirely in terms of the
vector-field $X^\mu$ directly.  The identification 
$X^\mu \equiv A^\mu - \partial^\mu \pi/m_X$
is exact, in the sense that renormalization does not disrupt
the size of the coefficient of $\partial^\mu \pi/m_X$
relative to $A^\mu$.  This follows from ensuring that 
the gauge transformations of $A^\mu$ and $\pi$ leave the combination
$A^\mu - \partial^\mu \pi/m_X$ invariant.

The purpose of introducing the ``fake'' gauge invariance is
to more easily uncover the role of the longitudinal
polarization of $X^\mu$, namely $X_L^\mu$, which for a suitable
choice of gauge, can be fully captured by the interactions of the
the scalar field $\pi$.  Hence, we will refer to $\pi$ as the
``longitudinal component'' synonymously with $X_L^\mu$,
though we emphasize that this identification is only strictly
true in Landau gauge, as we discuss below.

\subsection{BRST and $R_\xi$ gauge fixing}
\label{sec:BRST}

Before we discuss the gauge fixing of \eqnref{LApi} 
and applying the BRST to the St\"uckelberg formalism, 
we briefly review the general gauge-fixing and quantization procedure
using BRST~\cite{Becchi:1975nq,Tyutin:1975qk}.
The BRST transformations of the fields are equivalent to gauge transformations 
like those in \eqnref{Stucktransf} with infinitesimal gauge parameter
\eq{
\a(x) = \q \, \o(x) \,,
}
where $\q$ is an infinitesimal Grassmann constant and $\o$ is a real, Grassmann scalar field (ghost).
For the St\"uckelberg theory, we have the following BRST transformations of the fields:
\eqsl{BRSTtransf}{
\d_\q A &= \q \, \pd\o \,, \\
\d_\q \pi &= m_X \q \, \o \,, \\
\d_\q b &= 0 \,, \\
\d_\q \o &= 0 \,, \\
\d_\q \o^* &= \q b \,, \\
}
where $\o^*$ is a real, Grassmann scalar field (antighost) and $b$ is a Nakanishi--Lautrup auxiliary field \cite{Nakanishi:1966zz, Lautrup:1967zz}.
The action of a BRST operator $\mbf{s}$ on a field $\varf$ is defined in terms of the infinitesimal BRST transformation of a field $\varf$ by
\eq{
\d_\q \varf = \q \, \mbf{s}\varf \,.
}
For a product of fields,
\eq{
\d_\q (\varf_1 \varf_2) = (\d_\q \varf_1) \varf_2 + \varf_1 (\d_\q \varf_2) = \q \br{ (\mbf{s}\varf_1) \varf_2 \pm \varf_1 (\mbf{s}\varf_2) } \,,
}
where $\pm$ for whether $\varf_1$ is bosonic or fermionic; 
i.e., $\mbf{s}$ can be viewed as a fermionic operator. 
Using the transformations in \eqnref{BRSTtransf}, the gauge-fixing part of the Lagrangian can be written as \cite{Delbourgo:1987np}
\eql{Lgfb}{
\mcal{L}\ld{gf} = \mbf{s} \br{ \o^* \pr{ \mcal{G} + \frac{\xi}2 b } } = -\o^* \pr{ \mbf{s} \mcal{G} } + b \mcal{G} + \frac{\xi}2 b^2 \,,
}
where $\mathcal{G}[A,\pi]$ is a gauge-fixing functional.
Since $b$ is an auxiliary field and does not propagate, we can eliminate it using its EOM, 
yielding an alternate form for \eqnref{Lgfb},
\eql{Lgf}{
\mcal{L}\ld{gf} = -\o^* \pr{ \mbf{s} \mcal{G} } - \frac1{2\xi} \mcal{G}^2 \,.
}
The $R_\xi$-like class of gauge-fixing choices is obtained by setting
\eqsl{Rxi}{
\mcal{G}_\xi &= \partial_\mu A^\mu + \xi m_X \pi \,.
}
The general $R_\xi$-gauge Lagrangian is the sum of \eqnref{LApi} and the gauge-fixing terms,
\eqsl{LRxi}{
\mcal{L}_\xi &= \mcal{L}_g + \left. \mcal{L}\ld{gf} \right|_{\mcal{G}_\xi} \\
&= - \frac{1}{4} F_{A,\mu\nu} F_A^{\mu\nu} + \frac{1}{2} m_X^2 \left( A_\mu - \frac{\partial_\mu \pi}{m_X} \right)^2
- \frac{1}{2\xi}(\pd_\mu A^\mu + \xi m_X \pi)^2 - \omega^*(\partial^2 +\xi m_X^2)\,\omega \\
&= -\frac14 F_{A,\mu\nu} F_A^{\mu\nu}  -\frac1{2\xi}(\partial_\mu A^\mu)^2 + \frac 12 m_X A_\mu A^\mu + \frac12 \partial_\mu \pi \partial^\mu \pi - \frac12 \xi m_X^2 \pi^2 
- \omega^*(\partial^2 + \xi m_X^2)\,\omega \,,
}
which explicitly exhibits the decoupling of $A^\mu, \partial^\mu \pi$.%
\footnote{Using $R_\xi$ gauge fixing, the ghosts decouple in Abelian gauge theories because the ghost kinetic term involves only partial derivatives (in Yang--Mills theories, these become covariant derivatives in the adjoint representation). 
Hence, we omit them from the Lagrangian for the remainder of the paper.
}

From this, we see that the Proca Lagrangian corresponds to the choice $\xi \to 0$, where the second term in the last line of \eqnref{LRxi} decouples and $\pi$ becomes a free, massless scalar field.
The St\"uckelberg Lagrangian is obtained from the choice of St\"uckelberg--Feynman gauge $\xi = 1$,
\eql{LStuck}{
\mathcal{L}\ld{St} = - \frac{1}{4} F_{A,\mu\nu} F_A^{\mu\nu} 
+ \frac{1}{2} m_X^2 \left( A_\mu - \frac{\partial_\mu \pi}{m_X} \right)^2
- \frac12 (\pd^\mu A_\mu + m_X\pi)^2 \,.
}
Note that the first two terms in \eqnref{LStuck} are unchanged under the gauge transformation \eqnref{Stucktransf};
however, invariance of the last term requires $\pi$ to obey the EOM for a massive scalar field,
\eq{
\pr{\Box + m_X^2} \pi = 0 \,.
}

\subsection{Propagator in $R_\xi$ gauge}

The $R_\xi$ gauge fixing removes the mixing terms of the form
$A^\mu \partial_\mu \pi$ in the original St\"uckelberg Lagrangian
of \eqnref{LStuck}, 
leaving just the gauge-dependent two-point functions 
for $A^\mu$ and $\pi$. 
These have the standard $R_\xi$-gauge forms:
\eqsl{Api2pt}{
  \langle A^\mu(p) A^\nu(-p) \rangle &= \frac{-i}{p^2 - m_X^2} \left( g^{\mu\nu} - \frac{p^\mu p^\nu}{p^2 - \xi m_X^2} (1-\xi) \right) \,, \\
  \langle \pi(p) \pi(-p) \rangle &= \frac{i}{p^2 - \xi m_X^2} \,.
}
Using \eqnref{BRSTtransf} and the decomposition in \eqnref{XApi}, the BRST transformation of $X^\mu$ is
\eq{
\delta_\theta X^\mu = \delta_\theta A^\mu - \frac{1}{m_X} \partial^\mu \delta_\theta \pi
= \theta \partial^\mu \omega - \partial^\mu \left( \theta \omega \right) = 0 \,.
}
$X^\mu$ is annihilated by the BRST operator and corresponds to a
physical external state.
From \eqnref{Api2pt}, the $X^\mu$ two-point function can be reconstructed as
\eqsl{X2pt2}{
\langle X^\mu(p) X^\nu(-p) \rangle
&= \langle A^\mu(p) A^\nu(-p) \rangle
      + \frac{1}{m_X^2} (i p^\mu) (-i p^\nu)
      \langle \pi(p) \pi(-p) \rangle \\
&= \frac{-i}{p^2 - m_X^2}
      \left( g^{\mu\nu} - (1-\xi) \frac{p^\mu p^\nu}{p^2 - \xi m_X^2} \right)
      + \frac{i}{m_X^2} \frac{p^\mu p^\nu}{p^2 - \xi m_X^2}  \\
&= \frac{-i}{p^2 - m_X^2} 
      \left( g^{\mu\nu} - \frac{p^\mu p^\nu}{m_X^2} \right) \, ,
}
which agrees with \eqnref{X2pt1}. 
The absence of $\xi$-dependence demonstrates that the propagator for
the physical state $X^\mu$ is, unsurprisingly, itself
independent of the fake gauge symmetry.

\subsection{Current conservation}

The decomposition $X^\mu \equiv A^\mu - \partial^\mu\pi/m_X$
allows us to study St\"uckelberg theories using techniques familiar
from gauge theories. In fact, the fake gauge field $A^\mu$
has the same form as that of a massive gauge field arising
from a Higgsed $U(1)$ symmetry that is spontaneously broken
with mass $m_X = g v/2$.  However, for a St\"uckelberg vector field,
we know that only the combination $A^\mu - \partial^\mu\pi/m_X$
is physical and can represent an external state, while for a
gauge theory, the external state is of course just $A^\mu$. 
How do we reconcile this difference?

To understand when there is a distinction between
the St\"uckelberg vector field and a spontaneously broken
massive gauge field $A^\mu$, we examine the BRST current,
\begin{eqnarray}
J^\mu_{\rm BRST} &=& \sum_{{\rm field} \; \varf}
                     \frac{\delta L}{\delta_\theta \partial_\mu \varf} \delta_\theta \varf \,.
\end{eqnarray}
To keep things simple, consider a scenario in which the spin-one fields have interactions with a fermion current,
i.e., $g\,(A_\mu - \partial_\mu \pi/m_X) j^\mu_{\rm ferm}  \equiv g X_\mu j^\mu_{\rm ferm} $ for a St\"uckelberg vector field and $g A_\mu j^\mu_{\rm ferm}$ for a spontaneously broken, gauged $U(1)$ vector field. 

In the case where $A^\mu$ is a gauge field that is spontaneously broken, it is straightforward to show that the divergence of the BRST current is
\begin{eqnarray}
  \partial_\mu J^\mu_{\rm BRST}
  &=& - \omega \, \partial_\mu j^\mu_{\rm ferm} \qquad
      \mbox{(for a massive gauge field} \; A^\mu) \,.
\end{eqnarray}
Therefore, a conserved BRST charge requires the divergence of
the fermion current to vanish.

In contrast, when the same BRST transformations are applied to the
St\"uckelberg vector field, we obtain 
\begin{eqnarray}
\label{eqn:BRSTstuck}
  \qquad \qquad \partial_\mu J^\mu_{\rm BRST}
  &=& 0 \quad \mbox{(for a St\"uckelberg vector field} \; X^\mu = A^\mu - \partial^\mu \pi/m_X)\,.
\end{eqnarray}
A conserved BRST charge can always be formed since
the divergence of the BRST current vanishes independently
of the conservation of the fermion current.

Once we enforce a conserved current (in what follows, a fermionic
current), under the decomposition $X^\mu = A^\mu - \partial^\mu \pi/m_X$
the scalar field $\pi$ decouples from this interaction leaving
$X^\mu$ and $A^\mu$ indistinguishable.

\section{Tree-level Couplings of a St\"uckelberg Vector Field}
\label{sec:treelevelstuckelberg}

We now turn to considering the tree-level interactions
of a St\"uckelberg vector field $X^\mu$. 
As we have emphasized, $X^\mu$ does not transform under a gauge symmetry.  
Hence, interactions in the effective theory will be built from powers of $X^\mu$.
The goal in this section is to enumerate the possible renormalizable
tree-level interactions of $X^\mu$
and identify those that lead to scattering amplitudes that grow with
powers of $\sqrt{s}/m_X$.  These amplitudes
arise from couplings of the longitudinal mode $X^\mu_L$.
The absence of a (gauge) symmetry under which $X^\mu$ transforms
implies that its mass does not signal spontaneous symmetry breaking (SSB)
nor the existence of Goldstone bosons.
Nevertheless, the longitudinal mode, $X^\mu_L$, is physical.  
We now state the \textit{longitudinal equivalence theorem}: 
the leading interactions of the longitudinal mode can be characterized
either by working directly with $X^\mu_L$, or by using the
fake gauge invariance of \eqnref{Stucktransf}, 
choosing Landau gauge, and then associating $X^\mu_L$ with
the interactions of the derivatively coupled longitudinal scalar field $\pi$.
This is the St\"uckelberg analogue of the Goldstone boson equivalence theorem.

\subsection{The generalized Ward identity and the longitudinal equivalence theorem}
\label{sec:long}

In a theory with an exact $U(1)$ gauge symmetry, current
conservation leads to the Ward identity
\begin{eqnarray}
  k^\mu \mathcal{M}_\mu
  &=& 0 
\end{eqnarray}
for an arbitrary amplitude $\mathcal{M}$ in momentum space.
This implies that the longitudinal polarization of an external on-shell
gauge boson decouples.
For a spontaneously broken $U(1)$ gauge theory, 
in which a gauge field $A^\mu$ acquires a mass $m_X$, 
the longitudinal polarization of an external on-shell gauge boson has, 
of course, physical couplings.  Again using current conservation,
a generalized Ward identity
\begin{eqnarray}
  \frac{k^\mu}{m_X} \mathcal{M}_\mu(A) &=& i \mathcal{M}(G^0; \xi=0)
\label{eqn:AG0ward}
\end{eqnarray}
can be constructed that relates the momentum-contracted amplitude for an on-shell external gauge boson
$A^\mu$ with momentum $k^\mu$ with the same amplitude,
$\mathcal{M}(G^0; \xi=0)$, for the Goldstone boson in Landau gauge.
In the limit of large momentum $|\vec{k}| \gg m_X$,
$\epsilon_L^\mu(A) \simeq k^\mu/m_X$, giving the Goldstone boson
equivalence theorem
\begin{eqnarray}
  \epsilon_L^\mu(A) \mathcal{M}_\mu(A)
  \; \stackrel{|\vec{k}| \gg m_X}{\longrightarrow} \; 
  i \mathcal{M}(G^0; \xi=0)
\label{eqn:AG0generalized}
\end{eqnarray}
for a single on-shell, longitudinally polarized gauge boson.

For the massive St\"uckelberg vector boson, there is no
(generalized or other) Ward identity since there is no
gauge symmetry and thus no conserved local current
associated with $X^\mu$.
This means
\begin{eqnarray}
      \frac{k^\mu}{m_X} \mathcal{M}_\mu(X) \; \not= \; 0 \, .
\label{eqn:Xgeneralized}
\end{eqnarray}
At large momentum $|\vec{k}| \gg m_X$,
$\epsilon_L^\mu(X) \simeq k^\mu/m_X$, and so this is simply
a statement that the longitudinal mode of a St\"uckelberg vector
field couples with a strength of $k^\mu/m_X$.

What if we follow \secref{fakegauge} and \secref{BRST} and decompose 
the St\"uckelberg vector field into a fake gauge boson $A^\mu$ and 
scalar field $\pi$ and use the fake gauge invariance and $R_\xi$
gauge fixing to remove the $A^\mu \partial_\mu \pi$ mixing terms?
Here, the gauge redundancy of $A^\mu$ and $\pi$ implies that there is no 
gauge-independent identification of $X_L^\mu$ with $A_L^\mu$ and/or $\pi$.
Consider the two-point functions \eqnref{Api2pt} and \eqnref{X2pt2}.  
As we have discussed, the sum of the polarizations of $X^\mu$ is gauge independent.
We can match the sum of the polarizations of $X^\mu$
to that of a massive gauge field $A^\mu$ by going to
unitary gauge, $\xi \to \infty$.
In unitary gauge, $\pi$ does not play a dynamical role 
because $m_\pi^2 = \xi m_X^2 \to \infty$, and so 
$\epsilon_L^\mu(X) = \epsilon_L^\mu(A; \xi \to \infty)$.
By contrast, in Landau gauge ($\xi = 0$) the sum of the polarizations
of the two-point function of $A^\mu$ is purely transverse,
matching that of a massless gauge theory that has only two propagating DOF.  
Hence, in Landau gauge, the longitudinal polarization
$X^\mu_L$ is fully captured by $\partial^\mu \pi/m_X$.  
This is the same result found in a spontaneously broken gauge theory in Landau gauge,
where the longitudinal polarization of a massive gauge field is fully captured by 
$\partial^\mu G^0/m_X$ for the eaten Goldstone scalar field.  

Therefore, analogously to \eqnref{AG0generalized} for a spontaneously broken theory,
in Landau gauge at large momentum $|\vec{k}| \gg m_X$, we can identify
\begin{eqnarray}
  0 \; \not= \; 
  \epsilon_L^\mu(k) \mathcal{M}_\mu(X)
  \; \stackrel{|\vec{k}| \gg m_X}{\longrightarrow} \; 
  \frac{k^\mu}{m_X} \mathcal{M}_\mu(X) &=& i \mathcal{M}(\pi; \; \xi = 0) \, .
\label{eqn:XLgeneralizedLandau}
\end{eqnarray}
This is the \textit{longitudinal equivalence theorem}: the leading behavior
for on-shell, external $X_L^\mu$ interactions can be found by replacing
$X_L^\mu$ with $\partial^\mu \pi/m_X$. 
For St\"uckelberg theories, longitudinal equivalence arises as a consequence of the invariance of Green's functions under BRST transformations (Slavnov--Taylor identities) carried out on the $A_\mu - \partial_\mu \pi/m_X$ formulation. 
Following \eqnref{BRSTstuck}, BRST invariance holds for St\"uckelberg theories regardless of whether $A_\mu - \partial_\mu \pi/m_X$ couples to conserved currents. 
Goldstone equivalence in a Higgsed $U(1)$ theory can also be formulated from BRST invariance (assuming $\partial_\mu j^\mu_{\rm ferm} = 0$); 
however, it is more commonly derived using the generalized Ward identities from $U(1)$ gauge invariance (gauge fields coupling to conserved currents).%
\footnote{See refs.~\cite{Wulzer:2013mza, Cuomo:2019siu} for more details on the relation between the BRST Slavnov-Taylor identities and the generalized Ward identity in this regard.}
Moreover, we can also identify the leading behavior of the
off-shell two-point function \cite{Bagger:1989fc}, 
\begin{eqnarray}
  \langle X^\mu(k) X^\nu(-k) \rangle 
  \; \stackrel{k^2 \gg m_X^2}{\simeq} \; 
  \frac{k^\mu k^\nu}{m_X^2} \langle \pi(k) \pi(-k); \xi = 0 \rangle \, .
\label{eqn:XLpropagatorLandau}
\end{eqnarray}

The St\"uckelberg formalism makes clear that
the large-momentum behavior found by using
\eqnref{XLgeneralizedLandau} and \eqnref{XLpropagatorLandau}
yields nonrenormalizable interactions of the longitudinal mode
$\pi$ suppressed by powers of $m_X$.
Below, we will utilize these results in our discussions of the leading behavior
of interactions and scattering amplitudes at large momentum.
We note that Lagrangians involving $X^\mu$ do not necessarily contain interactions
of the longitudinal mode $\pi$.  For example, one special case is the
Proca Lagrangian \eqnref{LProca}
\begin{eqnarray}
  -\frac14 F_{X,\mu\nu} F_X^{\mu\nu} + \frac12 m_X^2 X_\mu X^\mu
  &\stackrel{k^2 \gg m_X^2}{\to}& \frac12 \partial_\mu \pi \partial^\mu \pi \,,
\end{eqnarray}
i.e., by the equivalence above, the Proca Lagrangian for a free massive St\"uckelberg 
vector field becomes the Lagrangian for a free massless scalar field $\pi$.

We now turn to considering interactions of $X^\mu$ with itself or with the SM, 
identifying those interactions that couple to the longitudinal mode, 
and discussing the consequences for the effective field theory.

\subsection{Conserved vector current}

Consider the interaction
\begin{eqnarray}
g_X X_\mu j_{\rm V}^\mu \,,
\label{eqn:conservedcurrent}
\end{eqnarray}
in which the St\"uckelberg vector field couples to a conserved vector
current $j_{\rm V}^\mu$ with strength $g_X$.  For the purposes of this section,
the current is assumed to be exactly conserved,
$\partial_\mu j_{\rm V}^\mu = 0$.
(The anomalous case that leads to one-loop couplings
will be discussed in detail in \secref{anomalous}.)
Using the equivalence $X^\mu \equiv A^\mu - \partial^\mu \pi/m_X$,
it is clear that
the longitudinal component $\pi$ decouples from the conserved vector
current, since under integration by parts (IBP)
\begin{eqnarray}
  \frac{1}{m_X} (\partial_\mu \pi) j_{\rm V}^\mu &\rightarrow&
  -\frac{\pi}{m_X} \partial_\mu j_{\rm V}^\mu \; \rightarrow \; 0 \, .
\end{eqnarray}
This is the famous example of a dark photon
kinetically mixed with electromagnetism, namely
$j_{\rm V}^\mu = j_{\rm em}^\mu$, with coupling strength $g_X = \epsilon \, e$ \cite{Holdom:1985ag}.  
This coupling is equivalent to a kinetically mixed 
St\"uckelberg field with the electromagnetic field strength
using the EOM $j_{\rm em}^\mu = \partial_\rho F_{\rm em}^{\rho\mu}$ and IBP\@. 
In the electroweak theory, $j_{\rm V}^\mu = j_Y^\mu$, 
with coupling strength $g_X = \epsilon g'/c_W$, where $c_W$ is the cosine of the Weinberg angle.
While $j_Y^\mu$ is no longer a pure vector current, it of course remains
anomaly-free.  (The couplings of $X$ to the axial vector part
of hypercharge will be discussed in the next section.)

While kinetic mixing
$\epsilon F_{X,\mu\nu} F_V^{\mu\nu}$ is equivalent to
$\epsilon g_V X_\mu j_V^\mu$, it is worth emphasizing that the inverse need not be true.
The St\"uckelberg vector field $X^\mu$ can be coupled to a
conserved current that is purely global and not gauged.
For example, in the SM the global current $j_{B-L}^\mu$ is
exactly conserved\footnote{The global $U(1)_{B-L}^3$ and
  $U(1)_{B-L} (\rm grav)^2$
  anomalies vanish in the presence of three right-handed neutrinos,
  though this is not critical to our argument.},
and so the interaction
\begin{eqnarray}
  g_X X_\mu j_{B-L}^\mu
\end{eqnarray}
can be written without explicitly gauging $B-L$.  This has fascinating
consequences when one imagines $X^\mu$ coupling to a linear combination
of both $j_{\rm em}^\mu$ and $j_{B-L}^\mu$ \cite{EsseiliKribs}. 
  
These statements also hold for St\"uckelberg vector fields coupled to 
currents of hidden (dark) fermions, which are commonly found in the literature. 
In this scenario, it is often assumed that $X^\mu$ is the gauge boson of a 
new $U(1)$ and that the hidden fermions are charged under this symmetry. 
However, provided the hidden current coupling to $X^\mu$ is 
vector-like and conserved, this need not be the case---the 
interaction is indistinguishable from a St\"uckelberg vector field 
coupled to a global (hidden fermion) current.

\subsection{Axial-vector current} 
\label{sec:axialcurrent}

Next, consider an interaction of $X^\mu$ with an axial current,
\eql{axialvector}{
g_X X_\mu j_{\rm A}^\mu \, .
}
Unlike the case of the global vector current, the global axial-vector
current is not, in general, conserved already at tree-level.  This is simply
because an axial-vector current 
is explicitly violated by fermion masses
(within the SM or beyond).

The consequences of the axial-vector current violation by fermion mass is
most easily seen by focusing on the longitudinal component of
the St\"uckelberg vector field, $X_L^\mu$, or equivalently
$-\partial^\mu \pi/m_X$ following \eqnref{XApi}.
For an axial current of fermions
$j^\mu_{\rm A} = \overline{f} \gamma^\mu \gamma_5 f$, 
the $\pi$ field is derivatively coupled,
so the longitudinal part of \eqnref{axialvector} is
\begin{align}
g_X X_{L, \mu} j_{\rm A}^\mu
\rightarrow -\frac{g_X}{m_X} \pd_\mu\pi\,(\bar f \gamma^\mu \gamma_5 f) 
\label{eqn:longitudinalaxialcoupling}
\end{align} 

\begin{figure}[t!]
\begin{center}
\includegraphics[bb = 200 360 500 450]{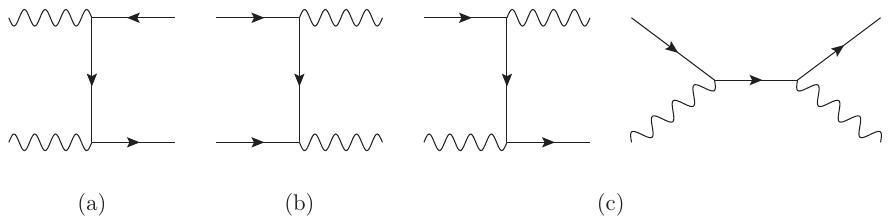}
\end{center}
\caption{Diagrams for 2--2 scattering amplitudes with two fermions
  and two gauge bosons:
  (a) $X X \to f\bar{f}$; 
  (b) $f\bar{f} \to X X$; 
  (c) $f X \to f X$.  [We have omitted the $u$-channel diagrams for (a),~(b).] 
  When $\partial_\mu j_{\rm A}^\mu \not= 0$ 
  due to the explicit violation of the global axial current by the
  fermion mass, the amplitudes for the
  longitudinally-polarized $X$ field grow with energy
  proportional to $m_f \sqrt{s}/m_X^2$.
}
\label{fig:XLXLff}
\end{figure}

We can use this result to illustrate the high energy behavior of $X^\mu$
in several scattering processes that have axial-vector couplings
including 
$f\overline{f} \ra X X$, 
$X X \ra f\overline{f}$, 
and $f X \ra f X$ as shown in Fig.~\ref{fig:XLXLff}.

The full expression for the scattering amplitude
follows by using the vector-boson polarization tensor
for the external boson $X^\mu$.
Since we are interested in the high-energy behavior of
the amplitude, we can focus on just the longitudinal part
using \eqnref{longitudinalaxialcoupling}.~\footnote{We can move the derivative off of the $\pi$ and onto the axial current via integration by parts, then enact the EOM by a field redefiniton $f \to  (1 -i\,g_X\, \pi\,\gamma_5/m_X - g^2_X\,\pi^2/(2m^2_X) )f$. This transforms the interactions in the Lagrangian to: $ \frac{2\,i\, m_f\, g_X}{m_X}\,\pi \,\bar f \gamma_5 f  - \frac{2\,m_f\, g^2_X}{m^2_X}\,\pi^2 \bar f f $. Notice that we have to go beyond the naive use of the EOM within the Lagrangian and work to $\mathcal O(1/m^2_X)$ as the higher order term contributes to the $X_L X_L \to \bar f f$ amplitude with the same strength as two insertions of the lower order vertex. After the redefinition, all interactions are proportional to the fermion mass.}

Using this effective interaction, and taking the limit of
$s \gg m_X^2, m_f^2$ with $t$ fixed in the amplitude squared,
we find that $X_L X_L \to \bar f f$ is 
\eq{
{|\mathcal M|^2 = \frac{128\, g^4_X\,m^2_f\,s\, t}{m^4_X\,(t- m^2_f)} + \cdots}
}
where $\cdots$ stands for terms with subdominant energy growth.
We see that the amplitude grows with energy
proportional to $m_f \sqrt{s}/m_X^2$.

We can obtain a crude estimate of the scale at which perturbative unitarity is
violated by setting $|t| \gg m^2_f$ and setting $|\mathcal M|^2 = 1$ (or by integrating over the full range of scattering angles and setting $\int d\cos\theta |\mathcal M|^2 = 1$),
\begin{eqnarray}
\sqrt{s_{\rm{max}}} &\sim& \frac 1 {8\sqrt 2}\frac{m^2_X}{g^2_X\,m_f} \,.
\end{eqnarray}
The effective theory for a St\"uckelberg vector field has a cutoff
scale that is parametrically above $m_X$ only when $m_f \ll m_X$.
This is fully equivalent to the Appelquist--Chanowitz bound
on scattering amplitudes involving longitudinal electroweak gauge
bosons and SM fermions when the Higgs is decoupled from the SM
\cite{Appelquist:1987cf}.

If instead the fermions are much heavier than the scattering
energy, $m_f^2 \gg s \gg m_X^2$, the fermions can be integrated out,
generating an effective $(X_\mu X^\mu)^2$ quartic interaction at one-loop order
that will also lead to amplitudes that grow with energy.
This is investigated below in \secref{quartic}.

The coupling of $X^\mu$ to an axial-vector current is equivalent
to a dimension-4 Higgs-derivative interaction with $X^\mu$
\eql{HDHX}{
  i H^\dagger \overleftrightarrow D_\mu H X^\mu \, ,
}
where we remind the reader that 
$H^\dagger \overleftrightarrow D_\mu H = H^\dagger (D_\mu H) - (D_\mu H^\dagger) H$
is a SM gauge singlet with fully
contracted $SU(2)_L \times U(1)_Y$ indices.
Focusing on the longitudinal part,
\eq{
i H^\dagger \overleftrightarrow D_\mu H X_L^\mu \to
-i H^\dag \overleftrightarrow D_\mu H \frac{\pd^\mu \pi}{m_X} \, .
}
Using IBP, the longitudinal coupling becomes
\eql{higgscurrent1}{
  i \frac{\pi}{m_X} \, \pd^\mu (H^{\dag}\overleftrightarrow{D}_\mu H) =
       i \frac{\pi}{m_X} \, \left[ H^{\dag}D^2H - (D^2H^{\dag})H \right] \, .
}
In the last line, we are free to promote the partial derivative to a
covariant derivative since the additional SM vector boson terms needed
to covariantize the left-hand side of \eqnref{higgscurrent1} vanish under $\overleftrightarrow{D}$.
Applying the EOM of the Higgs field, the Higgs mass and quartic
will cancel,
leaving just the $\pi$ coupling to a pseudoscalar current
proportional to Yukawa couplings,
\begin{align}
\rightarrow -i \frac{\pi}{m_X} \, \pi (\bar f_L y_f\, f_R - \bar f_R\, y^{\dag}_f f_L) \frac{(v + h)}{\sqrt 2} \, .
\end{align}
For the leptons and one type of quark, we can diagonalize the Yukawas so their entries are real and positive. 
In this case, 
\begin{align}
\rightarrow -i\,\, y_f \frac{(v + h)}{\sqrt 2}\, \frac{\pi}{m_X} \, (\bar f \gamma_5 f) \, .
\end{align}
We can convert this into an axial current by using the EOM for the fermions and IBP once more. 
Starting with \eqnref{higgscurrent1},
\begin{align}
& i\, \frac{\pi}{m_X} \pr{ -\bar e_R y^{\dag}_e\, (H^{\dag}L) + (\bar L H)\,y_e e_R + \cdots } \nonumber \\
= \ & i\, \frac{\pi}{m_X} \pr{ -\bar e_R i\slashed D e_R + \bar L i \slashed D L + \cdots } \nonumber \\
= \ & - \frac{\pd_\mu\pi}{2\, m_X} \pr{ \bar e_R \gamma^\mu e_R - \bar L \gamma^\mu L + \cdots } \nonumber \\
= \ & \frac{\pd_\mu\pi}{2\, m_X} (\bar f \gamma^\mu \gamma_5 f ) \,.
\label{eqn:HDHXmanip}
\end{align}
Hence, the Higgs-derivative interaction can be rewritten as axial-vector 
couplings of the SM fermions with $X^\mu$, and thus have the
same energy growth in the amplitudes.

If we do not immediately focus on the longitudinal piece of $X_\mu$, the operator $i H^\dagger \overleftrightarrow D_\mu H X^\mu$ contains a mass mixing between the $X$ and $Z$ and appears to lead to longitudinally enhanced $h \to X\, Z$ decays (assuming light $X$). However, as shown in Ref.~\cite{Dror:2018wfl}, once the mass mixing is removed all purely bosonic, longitudinally enhanced interactions are eliminated. The only longitudinally enhanced amplitudes come from SM axial current couplings to $X$ that arise as a consequence of the $X-Z$ mixing, in agreement with our result in \eqnref{HDHXmanip}.
 
While $X_\mu \overline{f} \gamma^\mu \gamma_5 f$ and $i H^\dagger \overleftrightarrow D_\mu H X^\mu$ 
separately lead to amplitudes that grow with energy, a carefully chosen combination of the two terms will not. 
This is precisely what occurs for $X^{\mu}$ coupling to the axial part of 
the hypercharge current \cite{Dror:2018wfl}. 
Explicitly,
\eql{hyperch}{
X_{\mu} j^\mu_{{\rm A},Y} \rightarrow  i\sum_f \frac{y_f(v+h)}{\sqrt 2 m_X} \Big( (Y_{f_R} - Y_{f_L}) \pm Y_H \Big)\,\pi\, (\bar f \gamma_5 f)
}
after carrying out the manipulations in \eqnref{longitudinalaxialcoupling} to \eqnref{HDHXmanip} 
and focusing on the longitudinal piece of $X^\mu$. 
Here $Y_{f_L}, Y_{f_R}$ are the hypercharges for $f_L$ and $f_R$, respectively,
$Y_H$ is the Higgs hypercharge, 
and the $+$ $(-)$ sign holds for leptons and down-type quarks (up-type quarks). 
Inserting the hypercharges for SM matter, \eqnref{hyperch} vanishes.
Thus, $X_{\mu} j^\mu_{{\rm A},Y}$ does not induce any amplitudes that grow with energy.

\subsection{Higgs portal}

At the renormalizable level, there is one independent Higgs interaction with $X^\mu$,
\eq{
  \frac{1}{2} \lambda_2 |H|^2 X_\mu X^\mu  \,.
} 
Inserting the Higgs vacuum expectation value (vev), this leads to an 
additional contribution to the mass of St\"uckelberg vector field.
The shifted mass is
\eq{
\tilde{m}_X^2 = m_X^2 + \frac{\lambda_2 v^2}{2} \, .
}
The interactions of the longitudinal component are identified as
\eq{
  \frac{1}{2} \lambda_2 |H|^2 X_\mu X^\mu 
  \to \frac{1}{2 \tilde{m}_X^2} \lambda_2  |H|^2 (\partial_\mu\pi \partial^\mu \pi) \,.
}
This yields dimension-5 and dimension-6 interactions of the
longitudinal mode $\pi$ with the Higgs field
\eql{higgsportallongitudinal}{
  \frac{\lambda_2 (2 v h + h^2)}{2 \tilde{m}_X^2}
  (\partial_\mu \pi \partial^\mu \pi)
}
that lead to scattering amplitudes that grow with powers of
$\sqrt{s}/m_X$.
Explicitly, examining the process $XX \to hh$ and using
$|\mathcal M|^2 = 1$ as the criterion for the perturbative unitarity
limit, we find
$\sqrt{s_{\rm{max}}} \sim \sqrt{\frac 2 {\lambda_2}}\tilde m_X.$

If $X^\mu$ were to acquire its mass mostly through this interaction
(i.e., $\tilde{m}_X^2 \simeq \lambda_2 v^2/2$), the strength of the
coupling $\lambda_2$ cancels out in \eqnref{higgsportallongitudinal}.
In this case, the St\"uckelberg vector boson amplitudes
grow with energy above the electroweak-breaking scale
independently of the mass of the St\"uckelberg vector boson.

Finally, we note that this operator is familiar from the scenario of a
$U(1)$ gauge field spontaneously broken by a complex scalar,
in which case $\lambda_2$ would be identified with $g^2$,
the square of the $U(1)$ gauge coupling.  This suggests that
$\lambda_2 < 0$ is highly suspect: in particular, the positivity of
$\lambda_2$ is mandatory in the case where the mass of
the St\"uckelberg field is obtained from this operator.

\subsection{Quartic self-interaction} 
\label{sec:quartic}

At the renormalizable level, there is one operator that leads
to a self-interaction of the St\"uckelberg vector field:
\eq{
  \frac{1}{4!} \lambda_4 (X_\mu X^\mu)^2 \,. 
}
For the longitudinal component this becomes
\eq{
  \frac{\lambda_4}{4! m_X^4} (\partial_\mu \pi \partial^\mu \pi)^2 \, .
}
In the presence of this quartic self-interaction,
the $2$--$2$ scattering amplitude with St\"uckelberg vector bosons
grows with energy as
\eq{
\mathcal{A}(X_L X_L \to X_L X_L) \sim \lambda_4 \frac{s^2}{m_X^4}
}
due to the couplings of the longitudinal mode.  The $s^2/m_X^4$
growth of the four-point amplitude is the same as that encountered
in the SM arising from (just) the four-point interaction of longitudinal
$W$ gauge bosons.  Of course, this energy growth 
is famously canceled in the SM by $Z$ and $h$ exchange diagrams.

The breakdown of the effective theory from this operator
can be obtained by performing a rough estimate of the
maximum allowed energy as in the previous subsection,
\begin{eqnarray}
\sqrt{s_{\rm max}} \lesssim \frac{m_X}{\lambda_4^{1/4}} \, .
\label{eqn:emax}
\end{eqnarray}
Separating $\sqrt{s_{\rm max}}$ and $m_X$ requires $\lambda_4 \ll 1$.%
\footnote{For vector-boson dark matter with $m_X \sim 10^{-5}$~eV 
\cite{Graham:2015rva} and requiring the cutoff scale to
be $\Lambda = M_{\rm Pl}$, we find an exceptionally small bound
on the coupling $\lambda_4 \lesssim 10^{-129}$.}

However, restricting to just the interactions of the normal dark photon model,
$(X_\mu X^{\mu})^2$ is not generated radiatively. 
The coupling $\lambda_4$ is multiplicatively renormalized and thus
technically natural if set to an exceptionally small number (including zero).  
It is well known that the sign of $\lambda_4$ must be positive to ensure
UV analyticity \cite{Adams:2006sv}. Moreover, $(X_\mu X^\mu)^2$ is generated -- even in the dark photon model -- if we appeal to the usual lore that quantum gravity generates all possible higher dimensional operators (suppressed by powers of $M_{\rm Pl}$). Specifically, we expect the operator
\begin{align}
\frac{(H^\dag H)(X_\mu X^\mu)^2}{M^2_{\rm Pl}},
\end{align}
which, below the scale of EWSB, leads to an effective quartic $\lambda_{4,{\rm eff}} \sim v^2/M^2_{\rm Pl}$. The growth of $\mathcal A(X_LX_L \to X_L X_L)$ is so rapid that even such a minuscule $\lambda_{4,{\rm eff}} \sim \mathcal O(10^{-32})$ can lead to perturbative unitarity violation at low scales (meaning well within the energy range we have probed experimentally) when $m_X$ is small. Explicitly, plugging this ``gravity-generated'' $\lambda_{4,{\rm eff}}$ into Eq.~\eqref{eqn:emax} we find $\sqrt{s_{\rm max}} \sim 10^8 \, m_X$.\footnote{Higher-dimensional operators formed from $X_\mu$ alone will also be generated by the same argument. Forming amplitudes from these, i.e., $\mathcal A(X_L X_L \to 4\,X_L)$ from $(X_\mu X^\mu)^3/M^2_{\rm Pl}$, we find a similar, but weaker bound on $\sqrt{s_{\rm max}}$.}

In the case of a Higgsed $U(1)$ theory in which the vector-boson
mass is acquired through SSB,
the energy growth of $X X \ra X X$ scattering is tamed by the
Higgs exchange diagram.  In the low-energy effective theory below
the mass of the Higgs (but above $m_X$), this interaction is generated
with a coefficient $\lambda_4 = 6 g^4 v^2/m_h^2$ where $m_X = g v/2$,
$m_h^2 = 2 \lambda_h v^2$, giving 
$\mathcal{A}(XX \to XX) \sim \frac{1}{\lambda_h} \frac{s^2}{v^4}$.
In other words, the scattering of vector bosons in a spontaneously broken 
$U(1)$ theory has an amplitude that grows with energy until the vev $v$, 
where the EFT must be supplemented by the Higgs boson.

\section{Coupling a St\"uckelberg vector field to an anomalous vector current}
\label{sec:anomalous}

Perhaps the most intriguing interaction that a St\"uckelberg vector
field could have is $X_\mu j^\mu_{\rm anom}$, a coupling to an  
anomalous current.  In this section we will mainly focus on coupling
to an anomalous \emph{vector} current, since we already showed
in \secref{axialcurrent} that a tree-level coupling to an
axial current generically leads to amplitudes that grow with
energy. 

For a gauge field,
$A_\mu j^\mu_{\rm anom}$ gauges what is a globally
anomalous $U(1)$ current associated with $j^{\mu}_{\rm anom}$.
In the presence of just one $U(1)$ gauge interaction, this leads
to the usual $U(1)^3$ anomaly.  When the fermions contributing to the
current $j^{\mu}_{\rm anom}$ also transform under other gauge symmetries,
such as the SM, this leads to the mixed anomalies $({\rm SM})^2 U(1)$.
The presence of the gauge anomalies leads to 
radiative corrections to the mass of the $U(1)$ gauge boson 
and to certain scattering amplitudes growing with energy
\cite{Preskill:1990fr,Craig:2019zkf}.

In \cite{Dror:2017nsg}, a detailed analysis of a light $U(1)$ gauge boson
coupled to an anomalous current was carried out. 
Their focus was on baryon number, which has the mixed anomalies 
$[U(1)_Y]^2 U(1)_B$ and $[SU(2)_L]^2 U(1)_B$.
The interaction $A_\mu j^{\mu}_{B}$ leads to couplings of the
longitudinal mode of $A_\mu$ with the (anomalous) baryon
current.  The consequences of this nonzero coupling emphasized in
\cite{Dror:2017nsg} are 
longitudinally enhanced interactions, including
$Z \rightarrow A \gamma$ and other anomaly-induced decays. 
A careful analysis of the loop functions leading to this decay
was carried out in \cite{Michaels:2020fzj}. 

But now there is a puzzle.  The St\"uckelberg vector field interaction
$X_\mu j^\mu_{\rm anom}$ appears to lead to an anomalous
fermion triangle loop, and yet, $X^\mu$ is not a gauge field.
There cannot be $U(1)^3$ or $({\rm SM})^2 U(1)$ mixed gauge anomalies
because there is no $U(1)$ gauge symmetry associated with $X^\mu$.

In this section, we resolve this puzzle and, in the course of our
analysis, find several consequences for theories
with a St\"uckelberg vector boson.
When we first introduce the fake gauge symmetry of \eqnref{Stucktransf},
the mystery seems to deepen further because now
$A_\mu$ would, in fact, appear to gauge an anomalous current.
We will see that the term $(\partial_\mu \pi/m_X) j^{\mu}_{\rm anom}$
precisely cancels the gauge anomaly that arises from $A_\mu j^{\mu}_{\rm anom}$. 
The mechanism responsible for canceling the anomaly can be understood essentially by IBP,
\eql{piibp}{
- \frac{\partial_\mu \pi}{m_X} j^{\mu}_{\rm anom}
\to \frac{\pi}{m_X} \partial_\mu j^{\mu}_{\rm anom} 
\propto \frac{\pi}{m_X} F_{\mu\nu} \tilde{F}^{\mu\nu} \,,
}
where $\wt{F}_{\mu\nu} \equiv \frac12 \e_{\a\b\mu\nu}F^{\a\b}$,
and we recognize that the partial derivative of the anomalous current $\partial_\mu j^{\mu}_{\rm anom}$
is proportional to $F_{\mu\nu} \wt{F}^{\mu\nu}$, the Chern--Pontryagin density.
The resulting dimension-5 interaction on the right-hand side of \eqnref{piibp} is referred to as the Peccei--Quinn term%
\footnote{This is also referred to as a 
  ``Green--Schwarz term'' in some of the literature, e.g.,
  \cite{Racioppi:2009yxa}.} 
(for any of the gauge symmetries of the SM, not just QCD).
When this term is combined with suitable Wess--Zumino terms%
\footnote{These are also referred to as
  ``generalized Chern--Simons terms'' in the literature, e.g.,
  \cite{Racioppi:2009yxa,Ekstedt:2017tbo}.}
(coupling a gauge/vector field to a Chern--Simons class%
\footnote{The Chern--Simons class for a non-Abelian gauge field is (the second term is zero for the Abelian case) 
\eq{
\O^\mu = \e^{\mu\nu\l\r} \pr{ A^a_\nu F^a_{\l\r} - \frac13 f^{abc} A^a_\nu A^b_\l A^c_\r }
\quad \Rightarrow \quad \pd_\mu \O^\mu = \frac12 \e^{\mu\nu\l\r} F^a_{\mu\nu} F^a_{\l\r} \,.
}
} 
\cite{Chu:1996fr, WeinbergVol1})
with appropriate choices of coefficients to restore gauge invariance, 
we will see that the Ward identities can be satisfied for 
all symmetries, verifying that $A_\mu$ does not have a gauge anomaly.%
\footnote{Coupling $A_\mu$ and $\pd_\mu \pi$, the two ``components" of $X_\mu$, separately to the Chern--Simons class for an unbroken gauge symmetry $\O_B^\mu$ yields
\eq{
A_\mu \O_B^\mu = A^\mu \e_{\mu\nu\l\r} B^\nu F_B^{\l\r} \,,
}
the dimension-4 Wess--Zumino term used to cancel the mixed anomaly, and
\eq{
\frac{\pd_\mu \pi}{m_X} \O_B^\mu = - \frac{\pi}{m_X} \pd_\mu \O_B^\mu = - \frac{\pi}{m_X} F_{B{\mu\nu}} \wt{F}_B^{\mu\nu} \,,
}
the dimension-5 Peccei--Quinn term.
As we will see in \secref{ward}, the four-dimensional Green--Schwarz mechanism combines these two types of terms to cancel anomalies.
}

We now turn to considering the coupling of a vector field to an 
anomalous symmetry current,
$j_{\rm anom}^\mu = \sum_{\j} q^\psi\, \j^\dag \bar{\s}^\mu \j$, 
where $q^\psi$ are the fermion charges under the symmetry. 
We wish to explicitly calculate the fermion loop attaching  
an external $A^\mu$ to two gauged vector bosons $B^\nu$ and $C^\rho$.  
Our discussion will apply to both a gauged vector field coupled
to an anomalous local symmetry current, as well as a St\"uckelberg
vector field $X^\mu$ coupled to an anomalous global symmetry current.
For the St\"uckelberg vector field, however, we will do this by
first decomposing $X^\mu = A^\mu - \partial^\mu \pi/m_X$, carrying
out the calculation of the contribution to the gauge anomaly
from $A^\mu$, and then add back in the contribution from
$\partial^\mu \pi/m_X$.

\begin{figure}[t!]
\begin{center}
\includegraphics[bb = 180 360 460 450]{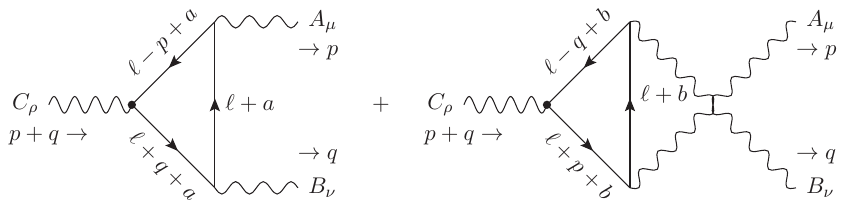}
\end{center}
\caption{Triangle diagrams responsible for the coupling of $A^\mu$
  (decomposed from $X^\mu = A^\mu - \partial^\mu \pi/m_X$)
  to two gauge bosons $B^\nu$ and $C^\rho$.
  We have labeled Lorentz indices and directions of four-momenta according
  to their use in the main text.
}
\label{fig:triangle}
\end{figure}

\subsection{Triple-gauge vertex from a single fermion loop}
\label{sec:vertex}

Consider the triangle diagrams that contribute to the anomaly 
with general vector bosons $A, B, C$ as shown in \figref{triangle}.
By power counting, their amplitudes are linearly divergent and thus not uniquely defined.
This can be encoded by including arbitrary four-momentum shifts $a$ and $b$ in the fermion loops in the left- and right-hand side diagrams, respectively. 
We will see that these arbitrary shifts are restricted by physical requirements, e.g., gauge invariance of either $B$ or $C$.

Our convention for the amplitude of the sum of the triangle diagrams in \figref{triangle} is
\eq{
\tilde{\D}^{\r\mu\nu}_{\{\mbf{r}\}} (p, q; m_\j; a, b) = g_C^{\mbf{r}_1} g_A^{\mbf{r}_2} g_B^{\mbf{r}_3} \, \tilde{\G}^{\r\mu\nu}_{\{\mbf{r}\}} (p, q; m_\j; a, b) \,,
}
where the indices $\mbf{r}_i \in \{\mrm{A},\mrm{V}\}$ indicate axial or vector couplings, 
respectively, of the boson with corresponding Lorentz superscript index in the same order, 
and $m_\j$ is the mass of the fermion $\j$ circulating in the loop. 
For now, all fermion charges have been subsumed into the couplings, 
so one should view $g^{\mbf{r}_{1,2,3}}_{A,B,C}$ as specific to the particular fermion in the loop, 
i.e., the interaction term in the Lagrangian for this fermion is
\eq{
\mcal{L}_{\rm int} = \bar{\j} \g^\mu \pr{g_C^{\rm V} - g_C^{\rm A} \g_5} \j \, C^{\mu} + \pr{C \to A, B} \,.
}

Focusing for example on the case $\mbf{r}_1 = \mrm{A}, \mbf{r}_2 = \mrm{V}, \mbf{r}_3 = \mrm{V}$, the amplitudes for the (coupling-stripped) triangle diagrams are:
\eqsl{vertexfuncAVV}{
\tilde{\G}_{\mrm{AVV}}^{\r\mu\nu} (p,q;m_\j;a,b) = \int_\ell \, \Tr\Bigg\{ &
\g_5 \g^\r \frac1{\slashed{\ell} + \slashed{a} - \slashed{p} - m_\j} 
\g^\mu \frac1{\slashed{\ell} + \slashed{a} - m_\j} 
\g^\nu \frac1{\slashed{\ell} + \slashed{a} + \slashed{q} - m_\j} \\
+ \ & \g_5 \g^\r \frac1{\slashed{\ell} + \slashed{b} - \slashed{q} - m_\j} 
\g^\nu \frac1{\slashed{\ell} + \slashed{b} - m_\j} 
\g^\mu \frac1{\slashed{\ell} + \slashed{b} + \slashed{p} - m_\j}
\Bigg\} \,,
}
where
\eq{
\int_\ell \equiv \int \frac{d^d \ell}{(2\pi)^d} \,.
}
For the $\mrm{VAV}$ and $\mrm{VVA}$ amplitudes, we move the $\g_5$ matrix in front of $\g^\mu$ or $\g^\nu$, respectively.
To avoid non-chiral anomalies, we set $b = -a$ \cite{Rosenberg:1962pp, WeinbergVol2, Dedes:2012me}.
In terms of the external momenta $p, q$, we can then express the arbitrary shift $a = z p + w q$ using two real parameters $z, w$.
The amplitude can be written in the Rosenberg parameterization as \cite{Rosenberg:1962pp, Michaels:2020fzj}
\eqsl{RosenbergVF}{
\tilde{\G}^{\r\mu\nu}_{\{\mbf{r}\}} & (p, q; z, w) =
\frac1{\pi^2} \bigg\{ G^1_{\{\mbf{r}\}} (p,q;w) \e^{\r\mu\nu ; p} + G^2_{\{\mbf{r}\}} (p,q;z) \e^{\r\mu\nu ; q} \\
+ \, & \Big( F_3(p,q) p^\mu + F_4(p,q) q^\mu \Big) \e^{\r\nu ; pq}
+ \Big( F_5(p,q) p^\nu + F_6(p,q) q^\nu \Big) \e^{\r\mu ; pq} \bigg\} \,,
}
where $\e^{\r\mu\nu ; q} \equiv \e^{\r\mu\nu \a} q_\a$ and we have made implicit the fermion mass dependence.
The form factors $F_i$ are finite and independent of $\{\mbf{r}\}$, whereas $G^1, G^2$ are dependent on the momentum shift $a$ (or the parameters $w,z$) as a consequence of the linear divergences of the triangle diagrams.

Full details of computing the form factors is given in \appref{Rosenberg}.
We quote here the final expressions for the $\mrm{AVV}$ and $\mrm{VAV}$ cases that we will use in the following sections.
Employing \eqnref{RosenbergFrelns} to eliminate $F_5$, we obtain for the $\mrm{AVV}$ and $\mrm{VAV}$ cases:
\eqsl{GAVVRosenberg}{
G^1_{\rm AVV} &= \frac14 (z+1) + p^2 \, F_3 - p \cdot q \, F_4 \,, \\
G^2_{\rm AVV} &= \frac14 (w-1) + q^2 \, F_6 + p \cdot q \, F_4 \,,
}
\eqsl{GVAVRosenberg}{
G^1_{\rm VAV} &= \frac14 (z+1) + p^2 \, F_3 - p \cdot q \, F_4 \,, \\
G^2_{\rm VAV} &= \frac14 (w-1) + q^2 \, F_6 + p \cdot q \, F_4  - m_\j^2 C_0(m_\j^2) \,.
}

\subsection{Momentum-contracted vertex functions}
\label{sec:mcvf}

Now that we have established how the triple-gauge vertex can be manipulated into purely finite terms -- 
form factors $F_{3\ldots6}$ plus the momentum-shift parameters $w,z$ -- 
we turn to its phenomenological consequences. 
The most interesting quantity is not the triple-gauge vertex itself, 
but what happens when the triple-gauge vertex is contracted with a longitudinally polarized $A, B,$ or $C$:
as explained in \secref{long}, the longitudinal polarizations are proportional to momenta in the large-momentum limit,
and these can lead to scattering or decay amplitudes that grow with energy.
The relevant quantities are the momentum-contracted vertex functions (MCVF)
\begin{align}
(p+q)_\r \, \tilde{\D}^{\r\mu\nu},\, p_\mu \, \tilde{\D}^{\r\mu\nu},\, q_\nu \, \tilde{\D}^{\r\mu\nu},
\end{align}
which are exactly the quantities we calculated in \appref{Rosenberg} to eliminate $G^{1,2}_{\{\mbf{r}\}}$.

In fact, the MCVF are the starting points for the calculation of the Ward identities for this vertex, 
e.g., for $A$ this is $p_\mu \mathcal{M}^\mu(A) = p_\mu \tilde{\Delta}^{\rho\mu\nu}$.
For a vertex that respects all of the symmetries, 
$(p+q)_\r \, \tilde{\D}^{\r\mu\nu} = p_\mu \, \tilde{\D}^{\r\mu\nu} = q_\nu \, \tilde{\D}^{\r\mu\nu} = 0$,
while for anomalous fermion content, one or more of these Ward identities is nonvanishing.
Contracting the momentum of a massive gauge boson with the vertex function
also yields a nonzero result, hence the Ward identity is also not satisfied.
However, as we discussed in \secref{long}, one can construct
a generalized Ward identity for a massive gauge boson
that relates the MCVF of the massive gauge boson with that having the massive gauge bosons swapped with
the Goldstone boson (for a spontaneously broken gauge symmetry) 
or the longitudinal mode (for a St\"uckelberg vector field).

Employing the procedure described in the previous subsection, we can compute
$(p+q)_\r \, \tilde{\D}^{\r\mu\nu}$, $p_\mu \, \tilde{\D}^{\r\mu\nu}$, $q_\nu \, \tilde{\D}^{\r\mu\nu}$ 
for $C, A, B$, respectively, with arbitrary combination of V, A couplings to the fermions in the loop.  
For the remainder of the paper, however, we will make the simplification that one of the vector fields, 
which we take (without loss of generality) to be $B$, has purely vectorial couplings. 
This is because the phenomenological examples we will examine in \secref{pheno} all share this property.  
The MCVF simplify to \cite{Michaels:2020fzj}:
\eqsl{MCVF}{
(p+q)_\r \, \tilde{\D}^{\r\mu\nu} &= \frac{g^{\rm V}_B}{4\pi^2} \e^{\mu\nu ; pq} \crl{ (w-z) \pr{g_C^{\rm V} g_A^{\rm A} + g_C^{\rm A} g_A^{\rm V}} + 4m_\j^2 C_0(m_\j^2) \cdot g_C^{\rm A} g_A^{\rm V} } \,, \\
-p_\mu \, \tilde{\D}^{\r\mu\nu} &= \frac{g^{\rm V}_B}{4\pi^2} \e^{\r\nu ; pq} \crl{ (w-1) \pr{g_C^{\rm V} g_A^{\rm A} + g_C^{\rm A} g_A^{\rm V}} - 4m_\j^2 C_0(m_\j^2) \cdot g_C^{\rm V} g_A^{\rm A} } \,, \\
-q_\nu \, \tilde{\D}^{\r\mu\nu} &= \frac{g^{\rm V}_B}{4\pi^2} \e^{\r\mu ; pq} \crl{ (z+1) \pr{g_C^{\rm V} g_A^{\rm A} + g_C^{\rm A} g_A^{\rm V}} } \,.
}
where
\eq{
\tilde{\D}^{\r\mu\nu} = \sum_{\mbf{r}_1, \mbf{r}_2 \in \{\mrm{A}, \mrm{V}\}} \tilde{\D}^{\r\mu\nu}_{\mbf{r}_1 \mbf{r}_2 \mrm{V}} \,,
}
and $C_0$ is a special case of the three-point Passarino--Veltman scalar function
\eql{C0}{
C_0(m_\j^2) = C_0(q^2, (p+q)^2, p^2; m_\j, m_\j, m_\j)
= - \int_0^1 dx \int_0^{1-x} dy \ \D^{-1} \,,
}
with $\D$ from \eqnref{Delta}.
Two relevant limits are 
\eqsl{C0limits}{
\lim_{m_\j^2 \to \infty} m_\j^2 C_0(m_\j^2) &\to -\frac12 \,, \\
\lim_{m_\j^2 \to 0} m_\j^2 C_0(m_\j^2) &\to 0 \,.
}
More precisely, these are limits of $m_\j^2$ with respect to the other scales $p^2, (p+q)^2, q^2$ that appear in \eqnref{Delta}.

In a theory with a fermion content that is nonanomalous, obviously
all of the MCVF in \eqnref{MCVF} vanish independently
of the presence or absence of masses for the vector bosons.
When the fermion content is anomalous, i.e.,
with respect to $A$ and/or $C$ (recall that we take $B$ to couple vectorially), 
the MCVF are not uniquely determined due to the
freedom to choose the coefficients of the most general momentum shift 
$a = zp + wq$ in the vertex function.
This allows for several possibilities. 
One possible choice of coefficients results in all three MCVF being equal,
\begin{align}
(p+q)_\rho \tilde\Delta^{\mu\nu\rho} = p_\mu \tilde\Delta^{\mu\nu\rho} =  q_\nu \tilde\Delta^{\mu\nu\rho} \ne 0 \, ,
\end{align}
a configuration referred to as the ``consistent anomaly''
\cite{Bardeen:1984pm,Hill:2006ei,Hill:2006ej}.
This choice is convenient from an EFT perspective:
we view the contributions to the gauge anomaly as arising from
the SM plus a contact term that, for instance,
arises from some heavy fermions that maintain anomaly cancellation.
In the consistent picture, all gauge symmetries are violated,
so integrating out UV physics can generate gauge-violating operators.
Combining these gauge-violating operators with the SM loop
(also gauge-violating in the consistent picture) and choosing
its coefficient appropriately, we can cancel all anomalies.%
\footnote{In the covariant picture, discussed below,
  the issue in the EFT is that
  integrating out UV physics can only change the coefficients of SM terms,
  or generate new, higher-dimensional terms that respect the UV symmetries. 
  As there is no $B$- or $C$-invariant, $A$-violating term, there is
  no coefficient to change, and the power counting for higher-dimensional terms 
  will not work out correctly to cancel the anomaly.
  Therefore, working in the covariant picture requires doing calculations in the full UV theory,
  keeping both SM and UV physics and not taking the low-energy limit of SM + effective operators.}

A second possibility is to utilize momentum shifts such that the
anomaly resides in only a single gauge interaction, the so-called
``covariant anomaly'' \cite{Hill:2006ei,Hill:2006ej}. 
Gauge-variant Wess--Zumino effective operators of the form 
$\epsilon^{\mu\nu\rho\sigma}A_\mu C_\nu F_{B,\rho\sigma}$
can be added to the Lagrangian to shift from the consistent
to covariant picture. 
This approach is often employed for calculations with two
gauge bosons, $B,C$, with anomaly-free couplings and one
(massive) gauge boson, $A$, that has anomalous couplings.  
By taking $w = z = -1$ in \eqnref{MCVF},
the terms that are independent of fermion mass appear only
in the MCVF for $A$ 
\begin{eqnarray}
  \label{eqn:MCVFCovariant}
  (p+q)_\r \, \tilde{\D}^{\r\mu\nu} &=& \frac{g^{\rm V}_B}{\pi^2} \e^{\mu\nu ; pq} m_{\j}^2 C_0(m_{\j}^2) \cdot g_C^{\rm A} g_A^{\rm V}  \,, \nonumber \\
  -p_\mu \, \tilde{\D}^{\r\mu\nu} &=& - \frac{g^{\rm V}_B}{2\pi^2} \e^{\rho\nu ; pq}
   \crl{ \pr{g_C^{\rm V} g_A^{\rm A} + g_C^{\rm A} g_A^{\rm V}} + 2 m_{\j}^2 C_0(m_{\j}^2) \cdot g_C^{\rm V} g_A^{\rm A} } \,, \\
  -q_\nu \, \tilde{\D}^{\r\mu\nu} &=& 0 \,. \nonumber
\end{eqnarray}
Notice that the fermion mass-dependent terms in the first two
expressions above come with different coupling structures: 
if the fermions have purely axial couplings to
$A$ (VAV structure),
the mass-dependent term vanishes from the first line, 
while if the couplings to $C$ are purely axial
(AVV structure),
the mass-dependent term vanishes from the second line.

\subsection{Anomaly cancellation, Ward identities, and $\pi$}
\label{sec:ward}

We are now in a position to clarify the role that the longitudinal mode
$\pi$ plays in anomaly cancellation.  Consider a theory
with massless fermions in which the vector field $A^\mu$ has anomalous couplings.  
For a single massless fermion $\j$, the MCVF become
\begin{eqnarray}
  \label{eqn:MCVFCovariantMassless}
  (p+q)_\r \, \tilde{\D}^{\r\mu\nu} &=& 0  \,, \nonumber \\
  -p_\mu \, \tilde{\D}^{\r\mu\nu} &=& - \frac{g_C g_X g_B q_B^{\psi}}{2\pi^2} \e^{\rho\nu ; pq}
   \pr{ q_{C}^{\mrm{V} ,\psi} q_{X}^{\mrm{A},\psi} + q_{C}^{\mrm{A},\psi} q_X^{\mrm{V},\psi} } \,, \\
  -q_\nu \, \tilde{\D}^{\r\mu\nu} &=& 0 \,, \nonumber
\end{eqnarray}
where from \eqnref{C0limits}, $m_\j^2 C_0(m_\j^2) \to 0$ in the massless fermion limit.  
Here, we have also separated the coupling constants $g_{X,B,C}$ from 
the individual fermion charges $q_{X,B,C}^\psi$ by writing
\begin{align}
  g_A^{\rm V} &= g_X \, q_X^{\mrm{V},\psi} \,, \, & \,
  g_A^{\rm A} &= g_X \, q_X^{\mrm{A},\psi} \, \nl
  g_B^{\rm V} &= g_B \, q_B^\psi \,, \, & \,
  g_B^{\rm A} &= 0 \,, \\
  g_C^{\rm V} &= g_C \, q_C^{\mrm{V},\psi} \,, \, & \,
  g_C^{\rm A} &= g_C \, q_C^{\mrm{A},\psi} \,. \nonumber
\end{align}
In this limit, the only nonvanishing MCVF is the one involving the $A^\mu$.
If we sum over several massless fermions, this becomes
\begin{eqnarray}
\label{eqn:MCVFanomcoef}
  -p_\mu \, \tilde{\D}^{\r\mu\nu} &=& - \mathcal{A}_X \frac{g_C g_X g_B}{2\pi^2} \e^{\rho\nu ; pq}
\end{eqnarray}
in terms of the $A^\mu$ anomaly coefficient
\begin{eqnarray}
  \label{eqn:anomcoefsum}
  \mathcal{A}_X &\equiv& \sum_\psi\,q^{\mrm{V},\psi}_B (q_{C}^{\mrm{V},\psi} q_{X}^{\mrm{A},\psi} + q_{C}^{\mrm{A},\psi} q_X^{\mrm{V},\psi}) \, .
\end{eqnarray}

From the start of \secref{vertex} until now, we have focused solely
on the contribution to the MCVF from a vector field $A^\mu$.  
Aside from forming the MCVF in \eqnref{MCVFanomcoef} by contracting the momentum 
of $A^\mu$ onto the vertex, we have not specified whether $A^\mu$ is massive or massless.
In addition, there is no distinction between whether $A^\mu$
is a gauge field that gauges the fermion current to which it couples with strength $g_X$ 
or is in fact a St\"uckelberg vector field $X^\mu$ that couples to a global fermion current with strength $g_X$.  

Below, we identify three distinct cases.
\begin{itemize}
\item[1.] \textit{$A^\mu$ is a massless gauge field}:  
  In this case, \eqnref{MCVFanomcoef} manifestly violates the Ward identity, 
  and so either $A^\mu$ must acquire a mass or the theory contains 
  multiple massless fermions with charges chosen such that, 
  while the contribution from any single (Weyl) fermion is nonzero,
  the sum in \eqnref{anomcoefsum} vanishes.
\item[2.] \textit{$A^\mu$ represents $X^\mu$, the massive St\"uckelberg vector field}:  
  In this case, \eqnref{MCVFanomcoef} is the final result for the MCVF
  that connects $X^\mu$ with the gauge fields $C^\rho$ and $B^\nu$ 
  through a loop of massless fermions. There is no (generalized or other) Ward identity
  since there is no symmetry or conserved current associated with $X^\mu$.
\item[3.] \textit{$A^\mu$ represents a massive gauge field arising from a 
  spontaneously broken $U(1)$ gauge symmetry}: 
  This is the conventional case,
  which requires additional \emph{massive} fermions, ``anomalons'', to cancel the anomaly.  
  The presence of anomalons is the key distinction from case 2.
\end{itemize}
We now want to compare and contrast cases 2 and 3,
but we first need to resolve the puzzle of decomposing 
$X^\mu = A^\mu - \partial^\mu \pi/m_X$.  In this decomposition,
$A^\mu$ is a gauge field, and so $A_\mu j^\mu_{\rm anom}$ necessarily 
gauges the anomalous current $j^\mu_{\rm anom}$; 
however, $X_\mu j^\mu_{\rm anom}$ is simply an interaction of a vector field
with a globally anomalous current $j^\mu_{\rm anom}$.  
How can $A^\mu$ be anomalous under its gauge symmetry while $X^\mu$ has
nothing to do with a gauge symmetry or a gauge anomaly?

The resolution is found by considering the additional contribution from 
the scalar field $\pi$.  The Lagrangian contains
\begin{eqnarray}
-g_X \frac{\partial_\mu \pi}{m_X} j^\mu_{\rm anom}
      &=& g_X \frac{\pi}{m_X} \partial_\mu j^\mu_{\rm anom} \,,
\end{eqnarray}
where we have used IBP to get the right-hand side. 
The divergence of the anomalous current is given by 
\begin{eqnarray}
  g_X \partial_\mu j^\mu_{\rm anom} &=&
     \mathcal{A}_X \frac{g_C g_X g_B}{4 \pi^2} 
     F_{C,\mu\nu} \tilde{F}_B^{\mu\nu} \, ,
\end{eqnarray}
and so the scalar field contributes a dimension-5 Peccei--Quinn term
in the Lagrangian, 
\begin{eqnarray}
     \mathcal{A}_X \frac{g_C g_X g_B}{4 \pi^2} 
     \frac{\pi}{m_X} F_{C,\mu\nu} \tilde{F}_B^{\mu\nu} \, . 
\label{eqn:PQtermpi}
\end{eqnarray}
In momentum space, this interaction becomes
\begin{eqnarray}
  i m_X \tilde{\D}^{\rho\nu}(\pi)
     &=& \mathcal{A}_X \frac{g_C g_X g_B}{2 \pi^2} \epsilon^{\rho\nu;pq} \,,
\label{eqn:piMCVF}
\end{eqnarray}
namely a dimension-5 three-point vertex among 
$\pi$, $C_\rho$, and $B_\nu$ in the effective theory.
We can combine \eqnref{MCVFanomcoef} with \eqnref{piMCVF} as
\begin{eqnarray}
p_\mu \tilde{\D}^{\rho\mu\nu}(A) - i m_X \tilde{\D}^{\rho\nu}(\pi) &=& 0 \, .
\label{eqn:Award}
\end{eqnarray}
This is the generalized Ward identity from \secref{long} for $A^\mu$ applied to the fermion triangle diagram.
That is, so long as the dimension-5 Peccei--Quinn term
has the specific coefficient given in \eqnref{PQtermpi},
$A^\mu$ satisfies the generalized Ward identity.
The specific coefficient that is required is precisely the one
that permits the combination of the renormalizable $A_\mu j^\mu_{\rm anom}$ 
and the dimension-5 interaction $- \partial_\mu \pi j^\mu_{\rm anom}/m_X$
to be written as $X_\mu j^\mu_{\rm anom}$;
in other words, the combination of $A_\mu$ and $(\pd_\mu \pi)/m_X$
must maintain the fake gauge invariance.  
This is otherwise known as the four-dimensional Green--Schwarz anomaly cancellation mechanism  \cite{Green:1984sg,Preskill:1990fr,Anastasopoulos:2006cz,Coriano:2007fw,Kumar:2007zza}.  

Since $A^\mu$ as part of $X^\mu$ is not an external state, we remark that 
\eqnref{Award}, the generalized Ward identity, is not a statement about longitudinal equivalence.
Contracting an on-shell external $X^\mu$
with $\tilde{\D}^{\r\mu\nu}$ in the high-momentum limit $|\vec{k}| \gg m_X$ 
gives \eqnref{MCVFanomcoef}, which we can equivalently calculate using 
an external on-shell $\pi$ and the longitudinal equivalence theorem in 
\eqnref{XLgeneralizedLandau}.  That is, $A^\mu$ and $\pi$ ``conspire'' 
to satisfy the generalized Ward identity while there is no analogue of this
for $X^\mu$. 

Finally, it is interesting to compare and contrast what happens
in a theory with a massive Abelian gauge boson in which the
anomalous contribution is canceled by anomalons.
The general case, with arbitrary vector and axial couplings for
$A^\mu$ and $C^\rho$, can be worked out straightforwardly from
\eqnref{MCVF}.  For the purposes of this discussion,
however, we simply illustrate the similarities and differences
in the case where the massless fermions contributing to the
anomaly have purely vector interactions to $A^\mu$ and $B^\nu$ and
purely axial interactions to $C^\rho$, in which case \eqnref{anomcoefsum}
simplifies to
\begin{eqnarray}
\label{eqn:coeffspecial1}
  \mathcal{A}_X &=& \sum_\psi\,q^{\mrm{V},\psi}_B q_{C}^{\mrm{A},\psi} q_X^{\mrm{V},\psi} \, .
\end{eqnarray}
The massive anomalons have purely axial interactions to $A^\mu$ and
purely vector interactions to $B^\nu$ and $C^\rho$,
\begin{eqnarray}
\label{eqn:coeffspecial2}
  \mathcal{A}_X^{\rm anom} &=& \sum_\psi\,q^{\mrm{V},\psi}_B q_{C}^{\mrm{V},\psi} q_X^{\mrm{A},\psi} \, .
\end{eqnarray}
Anomaly cancellation requires that the sum of the charges of the anomalons
under the gauge symmetries satisfy
\begin{eqnarray}
\label{eqn:anomcancellationcondition}
  \mathcal{A}_X^{\rm anom} &=& - \mathcal{A}_X \, ,
\end{eqnarray}
such that \eqnref{anomcoefsum} vanishes.  

However, for massive anomalons, there are additional
contributions to the momentum-contracted vertex function
from the $C_0$ functions in \eqnref{MCVF}.
We further simplify this discussion by
taking all of the anomalons to have the same mass $m_\psi$.
For the specific choices in
\eqnref{coeffspecial1} and \eqnref{coeffspecial2},
the only nonzero MCVF is
\eqsl{sumanomalon}{
  -p_\mu \, \tilde{\D}^{\r\mu\nu}
      &= - \frac{g_C g_X g_B}{2\pi^2} \e^{\rho\nu ; pq}
\left[ \mathcal{A}_X - \mathcal{A}_X^{\rm anom} 
\left( 1 + 2m^2_\psi C_0(m_\psi^2) \right) \right] \\
      &= \mathcal{A}_X^{\rm anom}
          \frac{g_C g_X g_B}{\pi^2} \e^{\rho\nu ; pq}
          m^2_\psi C_0(m_\psi^2) \, , 
}
where we used the anomaly cancellation condition
\eqnref{anomcancellationcondition} to get the second line.
If the anomalons were massless, the right-hand side above would vanish using \eqnref{C0limits};
this is as expected since by definition the theory would then be anomaly-free  
and the Ward identities satisfied.  
If the anomalons are infinitely massive, the term in parentheses
on the first line multiplying $\mathcal{A}_A^{\rm anom}$ vanishes using \eqnref{C0limits},
leaving the right-hand side nonzero and equal to \eqnref{MCVFanomcoef},
i.e., back to where we started.

With nonzero anomalon masses, \eqnref{sumanomalon} does not vanish.
Following our discussion in \secref{long}, we can again construct a
generalized Ward identity such that
\begin{eqnarray}
p_\mu \mathcal{M}^\mu(A) - i m_X \mathcal{M}(G^0) &=& 0 \, .
\end{eqnarray}
where, for this discussion,
$G^0$ is the Goldstone boson absorbed to make $A^\mu$ massive.
Applying this to the MCVF for the fermion triangle diagram, 
\begin{eqnarray}
p_\mu \tilde{\D}^{\rho\mu\nu} - i m_X \tilde{\D}^{\rho\nu}(G^0) &=& 0 \, .
\end{eqnarray}
From this we can deduce the required interaction that the Goldstone boson
must have with the MCVF,
\begin{eqnarray}
  i \tilde{\D}^{\rho\nu}(G^0)
      &=& \mathcal{A}_X^{\rm anom}
          \frac{g_C g_X g_B}{\pi^2} \epsilon^{\rho\nu;pq}
          \frac{m^2_\psi}{m_X} C_0(m_\psi^2) \, .
\label{eqn:MCVFgoldstone}
\end{eqnarray}
Here, we finally see the key difference between the case of a
St\"uckelberg vector field and a spontaneously broken massive
Abelian gauge field.  In the specific example above,
the anomalons have axial interactions with $A^\mu$, implying
the anomalons are \emph{chiral} with respect to the gauge symmetry
associated with $A^\mu$.  The only way to give mass to these
chiral fermions without explicitly breaking the symmetry is to write
Yukawa interactions with the Higgs field whose vev
spontaneously breaks the gauge symmetry associated with $A^\mu$.
This means that, with conventional normalizations
$m_\psi = y_\psi v/\sqrt{2}$ and $m_X = g v/2$,
one power of the vev drops out in \eqnref{MCVFgoldstone}.
Hence we see that the generalized Ward identity can be satisfied with
\emph{renormalizable} Yukawa interactions of the Goldstone mode
with the fermions.  This key difference is what permits
a spontaneously broken gauge symmetry with anomalous fermion
content (and a separate set of anomalons with heavier masses)
to be at least possibly viable without a divergence in the UV leading to
a cutoff scale.  The caveat is that this requires Yukawa couplings 
to be perturbative (i.e., less than order one) in order to avoid Landau poles.

\section{Applications to baryon number} 
\label{sec:pheno}

We now consider specific cases where the St\"uckelberg vector field
$X^\mu$ couples to a
globally anomalous current in order to investigate the
phenomenological consequences.  One of the most interesting
possibilities is $X^\mu$ coupling to baryon number.
Baryon number is anomalous in the SM, 
but anomaly-free with respect to $SU(3)_c \times U(1)_{\rm em}$ below the electroweak scale.
Here, our focus is to investigate the observable consequences of the
longitudinal enhancements that occur in the presence of
$X_\mu j^\mu_B$, specifically three observables:
$Z \to X\gamma$, $f\bar{f} \to X\gamma$, 
and $Z\gamma \to Z\gamma$.  
These depend on the electroweak scale and disappear in the limit $v \rightarrow \infty$.  
We compare and contrast our results with those when baryon number
is gauged \cite{Dror:2017nsg,Michaels:2020fzj}, 
identifying the similarities and differences for the
case of a St\"uckelberg vector field. 
In the discussion below, we take all SM fermions to be massless;
however, it is straightforward to re-introduce SM fermion mass dependence 
(e.g.,~\cite{Michaels:2020fzj}).  In reality, only the top quark
significantly invalidates this assumption, causing the baryon anomaly
coefficient to be slightly smaller than what we have assumed below.

\subsection{Prelude: $Z \to A\gamma$ with gauged baryon number}
\label{sec:ZXgamgauge}

As a prelude to the results in subsequent sections, we want to
review the calculation of $Z \to A\gamma$, where
$A^\mu$ is the gauge field associated with gauged baryon number 
\cite{Nelson:1989fx,Carone:1994aa,FileviezPerez:2010gw};
we reserve $X^\mu$ to refer to the St\"uckelberg vector field.
However, we will use $m_X$, $g_X$, and $q_X$ to refer to the
mass, coupling, and charges of the (gauged or ungauged) vector field
coupled to the baryon current.

In the SM, the baryon current is anomalous with respect to
the mixed anomalies $U(1)_Y^2 U(1)_B$ and $SU(2)_L^2 U(1)_B$
in the specific combination \cite{Dror:2017nsg}
\begin{eqnarray}
  \partial_\mu j^\mu_B
  &=& \frac{\mathcal{A}_B}{8\pi^2}
      \left( g'^2 B_{\mu\nu} \tilde{B}^{\mu\nu}
             - g^2 W_{\mu\nu} \tilde{W}^{\mu\nu} \right) \, .
\end{eqnarray}
Here $\mathcal{A}_B$ is the anomaly coefficient
\begin{eqnarray}
\mathcal{A}_B &=& \sum_{f \in {\rm SM}} Q^f q_X^{\mrm{V},f} q_Z^{\mrm{A},f} \, ,
\label{eqn:Bsum}
\end{eqnarray}
where the sum is over all of the fermions $f$ in the SM with
electric charge $Q^f$, 
baryon number $q_B^f$, 
and axial coupling $q_Z^{\mrm{A},f} = T_3^f/2 = \pm 1/4$ to the $Z$.
This is equivalent to the anomaly coefficient for
just $SU(2)_L^2 U(1)_B$ or (the negative of) $U(1)_Y^2 U(1)_B$
since $U(1)_{\rm em}^2 U(1)_B$ vanishes.  
Three generations of massless SM fermions give $\mathcal{A}_B = 3/4$.

As we have learned from \secref{mcvf}, 
we are free to choose a set of Wess--Zumino terms such that the only nonzero MCVF is
\begin{eqnarray}
  - p_\mu \sum_f \tilde \Delta^{\rho\mu\nu}_{\rm SM}
  &=& - \mathcal{A}_B \frac{e g g_X}{2 \pi^2 c_W} \epsilon^{\rho\nu;p q} \,,
      \label{eqn:wardnonzerogaugeB}
\end{eqnarray}
following \eqnref{MCVF} with the specific choices $w=z=-1$.%
\footnote{Had we included nonzero SM fermion masses, the first line of \eqnref{MCVF} would also be nonzero.
Adding $m_Z$ times the $Z_{\rho}$ Goldstone contribution, a $\phi_Z  - A - \gamma$ vertex, to the first line would yield zero.}

Baryon number can be made anomaly-free by extending the SM with anomalons $\psi$ 
with charges such that, when they are included in the sum \eqnref{Bsum}, 
the net result is zero. 
For certain choices of their $SU(2)_L \times U(1)_Y$ charges, 
these anomalons can obtain masses independently of the electroweak vev and therefore 
can be much heavier than the SM fermions. 
The full result for the decay rate $Z \to A\g$ including both SM fermions and a set of massive anomalons 
was given in \cite{Michaels:2020fzj}. 
However, for our purposes, it is more convenient to separate the contributions to the triangle loop 
from the SM, \eqnref{wardnonzerogaugeB}, and the massive anomalons. 
Defining $\tilde\Delta^{\rho\mu\nu}_{\rm anom}$ as the contribution to the vertex function from the anomalons, 
and making the same choice $w = z = -1$, the additional contribution to the MCVF can again be easily obtained from \eqnref{MCVF}, 
\begin{eqnarray}
  - p_\mu \sum_\psi \tilde\Delta^{\rho\mu\nu}_{\rm anom}
  &=&  \mathcal{A}_B \frac{e g g_X}{2 \pi^2 c_W} \epsilon^{\rho\nu;p q}(1 + 2m^2_\psi C_0(m_\psi^2)) \,.
\label{eqn:baryonanomaloncontribution}
\end{eqnarray}
To cancel the anomaly and obtain mass without electroweak symmetry breaking,
these anomalons have pure vector couplings to $Z$ and pure axial couplings to $A$
such that 
\eq{
\sum_\psi Q^\psi q_X^{\mrm{A},\psi} q_Z^{\mrm{V},\psi} = -\mathcal{A}_B \,.
}
This is the same situation we encountered in \secref{ward}---the
anomalon mass only appears in the MCVF in \eqnref{baryonanomaloncontribution}.

If we were instead to take $m_\psi \to 0$ (and therefore degenerate with the SM),
the two sectors would cancel exactly, as required of an anomaly-free theory. 
For nonzero anomalon masses, the cancellation between the two sectors is inexact, leaving
\eq{
 - p_\mu \sum_{f, \psi} \tilde\Delta^{\rho\mu\nu}_{\rm tot}
  =  \mathcal{A}_B \frac{e g g_X}{\pi^2 c_W} \epsilon^{\rho\nu;p q} m^2_\psi C_0(m_\psi^2) \,,
}
as in \eqnref{sumanomalon}. 
It is interesting to consider anomalons that are much heavier than the $Z$ boson.%
\footnote{To play a role in the anomaly, the anomalons must receive some of their mass 
from the same SSB that gives mass to $A$ and therefore $m_\psi \sim y\, v_X$, 
where $m_X \sim g_X v_X$ and $y$ is some Yukawa coupling. 
A large hierarchy between the anomalons and $X$ requires taking $g_X \ll y$, with the validity of perturbation theory limiting $y_{\rm max} \sim 4\pi$. 
More discussion on the phenomenological implications of this ``maximum hierarchy" between $\psi$ and $X$ can be found in \cite{Michaels:2020fzj}.}
Then the right-hand side simplifies to
\eql{manomheavy}{
 - p_\mu \sum_{f, \psi} \tilde\Delta^{\rho\mu\nu}_{\rm tot}
  =  -\mathcal{A}_B \frac{e g g_X}{2 \pi^2 c_W} \epsilon^{\rho\nu;p q} \, ,
}
which up to corrections of $\mathcal{O}(m_Z^2/m_\psi^2)$ reduces to just the original 
SM-only contribution in \eqnref{wardnonzerogaugeB}.
Dividing both sides by $m_X$, the above equation becomes the
amplitude for $Z \to A_L \gamma$, where $A_L$ is the longitudinal polarization.
Squaring, we can convert this to a decay rate
(again, in the limit that the SM fields are massless and the
anomalons are infinitely heavy)
\begin{eqnarray}
  \Gamma( Z \to A \g ) \; \stackrel{m_X \ll m_Z}{\simeq} \; 
  \Gamma( Z \to A_L \g ) 
  &\simeq& \frac{3}{32\pi^2} \frac{\alpha_{\rm em}^2 \alpha_X}{c_W^2 s_W^2}
       \frac{m_Z^3}{m_X^2} \, ,
\label{eqn:ZXgamma_gauged}
\end{eqnarray}
where $s_W$ is the sine of the Weinberg angle and 
we have used $\mathcal{A}_B = 3/4$.

As emphasized in \cite{Dror:2017nsg}, the $m_Z^2/m_X^2$ longitudinal
enhancement implies the decay width is unbounded in the limit
$m_X \ll m_Z$.  For the effective theory to be valid,
$\Gamma( Z \to A \g ) < m_Z$, which implies a lower bound on $m_X$ of
\eql{mbcutoff}{
  m_X > \sqrt{6 \pi} \times \frac{e g g_X}{64 \pi^3 c_W} \times m_Z \, .
}
Up to an irrelevant numerical prefactor, this is the same bound 
obtained by Preskill \cite{Preskill:1990fr} for an anomalous gauge theory 
by requiring the divergent three-loop contribution to the 
(anomalous) gauge boson mass not exceed its bare mass.  
More precisely, Preskill derived an expression $\Lambda = \frac {64 \pi^3 c_W}{e g g_X} m_X$
for the cutoff scale $\Lambda$ of the effective theory 
that has the same scaling as \eqnref{mbcutoff} when we
reinterpret the cutoff scale $\Lambda$ to be $m_Z$.

What happens when $m_X$ is lowered below the bound given in \eqnref{mbcutoff}?  
In a theory with anomalons, it is no longer possible to take their mass 
$m_\j$ to be much larger than $m_Z$.  
Approximating the results in \cite{Michaels:2020fzj} in the limit 
$m_X \ll m_\j \ll m_Z$ (with massless SM fermions), we find
\eq{
  \Gamma( Z \to A \g ) \simeq
      \frac{3}{32 \pi^2} \frac{\alpha_{\rm em}^2 \alpha_X}{c_W^2 s_W^2}
      \times m_Z \times
      \frac{m_\psi^4}{m_X^2 m_Z^2} \log^4 \frac{m^2_\psi}{m_Z^2} \,,
}
where now the EFT requirement 
$\Gamma( Z \to A \g ) < m_Z$ implies the lower bound on 
$m_X$ is modified to
\eql{mzcutoffanomalons}{
  m_X > \sqrt{6 \pi} \times \frac{e g g_X}{64 \pi^3 c_W} \times
              m_\j \times \frac{m_\j}{m_Z} \log^2 \frac{m_\j^2}{m_Z^2} \, .
}
This implies that we can lower the mass for $m_X$ at the price of reducing the
anomalon masses below $m_Z$.  However, the additional suppression factor 
$m_\j/m_Z \log^2 m_\j^2/m_Z^2$ on the right-hand side in \eqnref{mzcutoffanomalons}
relative to the result in \eqnref{mbcutoff} implies that the separation between 
$m_X$ and $m_\j$ can become increasingly large as $m_\j$ is lowered below $m_Z$.  

\subsection{$Z \to X\gamma$ with global baryon number} 
\label{sec:ZXgammaglob}

Now we are in a position to evaluate $Z \ra X\gamma$ when $X$
is a St\"uckelberg vector field with coupling $g_X X_\mu j_B^\mu$ to the
global, anomalous baryon current of the SM\@.  
The contribution to the $Z_\rho-X_\mu-\gamma_\nu$ vertex coming from 
loops of SM fermions is identical to the gauged case in the last section. 
Therefore, the nonzero MCVF is 
\begin{eqnarray}
  - p_\mu \sum_f \tilde{\Delta}^{\rho\mu\nu}_{\rm SM}
  &=& - \frac{\mathcal{A}_B}{2 \pi^2}\frac{e g g_X}{c_W} \epsilon^{\rho\nu;p q}
      \label{eqn:wardnonzero} \,.
\end{eqnarray}
This is the total contribution since there are no anomalons present. 

Using this vertex to calculate $Z \to X \gamma$, we find
\begin{eqnarray}
  \Gamma( Z \to X \g ) \; \stackrel{m_X \ll m_Z}{\simeq} \; 
  \Gamma(Z \to X_L \g) 
  &\simeq& \frac{3}{32 \pi^2} \frac{\alpha_{\rm em}^2 \alpha_X}{c_W^2 s_W^2}
      \frac{m_Z^3}{m_X^2} \, ,
      \label{eqn:ZXgamma_GS}
\end{eqnarray}
exactly the same result as \eqnref{ZXgamma_gauged}, the case where
$U(1)_B$ is gauged and made anomaly-free via infinitely
heavy anomalons.  We remark that the same result could also have
been obtained using the longitudinal equivalence theorem
to relate $\epsilon_L^\mu \tilde{\Delta}^{\rho\mu\nu}(X)$
to $\tilde{\Delta}^{\rho\nu}(\pi)$ in Landau gauge
at large momentum $|\vec{k}| \gg m_X$.

Thus we see that the gauging of the would-be anomalous baryon number
symmetry is irrelevant to the presence of the physically observable
decay process $Z \ra X\gamma$.  It is the presence of the
\emph{global} baryon number anomaly that is essential for this decay to proceed. 
Said differently, our results show that the decay rate alone cannot 
differentiate between the scenarios of a gauge boson accompanied by heavy anomalons 
and a St\"uckelberg field coupled to a global current---a 
perspective emphasized in \cite{Dedes:2012me}.  

The presence of the Peccei--Quinn term, a dimension-5 operator in the
St\"uckelberg EFT, implies a UV cutoff that cannot be taken arbitrarily large.
Applying \eqnref{PQtermpi} to the specific case of the anomalous baryon current,
the dimension-5 operator is 
\begin{eqnarray}
\label{eq:PQterm}
  \mathcal{A}_B \frac{ e g g_X}{4 \pi^2 c_W} \frac{\pi}{m_X}
  F_{Z,\mu\nu} \tilde{F}_{\rm em}^{\mu\nu} \,,
\end{eqnarray}
and requiring the coefficient of this operator be less than $4 \pi$,
we obtain a cutoff scale of order 
\begin{eqnarray}
\sqrt{s_{\rm max}} \sim \frac{16 \pi^3 c_W m_X}{\mathcal{A}_B e g g_X} \,.
\end{eqnarray}
The existence of a cutoff scale is not surprising because we previously
discovered in \eqnref{mbcutoff} that we could not arbitrarily separate 
$m_X$ from $m_Z$ while allowing the decay rate 
$\Gamma(Z \ra X\gamma)$ to remain perturbative.
Both bounds scale similarly (up to numerical coefficients)
with couplings and mass.  What we see is that a St\"uckelberg 
vector field coupled to a globally anomalous current has a
nonrenormalizable interaction signaling the existence of
amplitudes that can grow with energy. 
This is explicitly seen in the decay rate $Z \ra X\gamma$, 
and as we will see below, also occurs for processes that have one or more
factors of $\tilde{\Delta}^{\r\mu\nu}$ with an odd number of axial couplings 
embedded in the amplitude.

For finite $m_\psi \gg m_Z$, there will be corrections in \eqnref{ZXgamma_gauged}
of $\mathcal O(m^2_Z/m^2_\psi)$ that are absent in \eqnref{ZXgamma_GS}.
It is tempting to think that these corrections would be observable
given sufficiently accurate measurements of $m_X$ and
$\Gamma(Z \to X \gamma)$. However, this is premature,
since in the case of a St\"uckelberg vector field,
there are additional higher-dimensional operators
suppressed by $\Lambda$ that can contribute to the decay process. 
Hence, in the absence of direct observations (on-shell production)
of anomalons and/or a Higgs boson, there is no way to
unambiguously determine whether the decay process
signals the existence of gauged baryon number, or instead, a
St\"uckelberg vector field coupled to global baryon number.

\subsection{$f\bar{f} \to X\g$} 
\label{sec:fftoXgamma}

Attaching the $Z^\rho$ leg of the $Z^\r-X^\mu-\gamma^\nu$ vertex to a fermion current, 
we can explore how the longitudinal enhancement of the vertex manifests in $f \bar f \to X \gamma$, 
where $f$ is a SM fermion. 
This calculation is interesting because it allows us to probe the triple-gauge vertex and its longitudinal enhancement at a wider range of energies than in $Z$ decay.  
In particular, we can consider limits such as $m_X^2 \ll s \ll m_Z^2$, where the $Z$ has been integrated out. 

\begin{figure}[t!]
\begin{center}
\includegraphics[bb = 190 360 430 450]{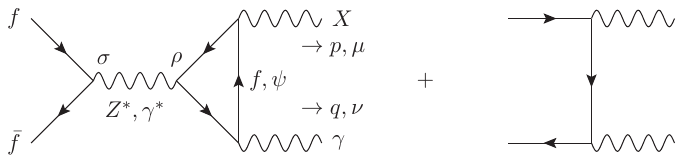}
\end{center}
\caption{Diagrams for $f\bar{f} \to X\g$, with $f$ an SM fermion: 
if only anomalons $\j$ couple to $X$, only the left diagram is relevant; 
otherwise, if $f$ also couples to $X$, there are the $t$- and $u$-channel diagrams as well 
(cross diagrams not shown).}
\label{fig:ffXgam}
\end{figure}

The diagrams for $f \bar f \to X \gamma$ are shown above in \figref{ffXgam}; 
an $s$-channel diagram proceeding through the triple-gauge vertex $\tilde \Delta^{\rho\mu\nu}$, plus $t$- and $u$-channel diagrams. 
The $t$- and $u$-channel diagrams involve only vectorial couplings and lead to the usual collinear divergences in the cross section. 
However, at least in the limit that the SM fermions are massless, they do not couple to the longitudinal part of $X$ and thus do not grow with $s$ (for a fixed scattering angle)%
\footnote{Additionally, the interference between the $t$- and $u$-channel diagrams and the $s$-channel diagram is zero.}. 
Therefore, we will ignore these diagrams and focus on the $s$-channel piece, deferring a more general calculation to \appref{fftoXgam}. 
Furthermore, we will focus on the $X_L$ piece of the amplitude, as this contains the leading dependence on $s$:
\eqsl{fftoXL}{
i\mathcal M(\bar f f \to X_L \gamma) = \frac{i g}{c_W}\frac{1}{s - m^2_Z} 
\bar{v}(k_2) \gamma^{\rho} \pr{ q^{\mrm{V}, f}_Z  - q^{\mrm{A},f}_Z \gamma_5 } u(k_1) 
\tilde \Delta^{\rho\mu\nu}_{\rm tot} \frac{p_\mu}{m_X} \e^*_\nu(q) \,,
}
where $q^{\mrm{V}, f}_Z \ (q^{\mrm{A}, f}_Z)$ are the vectorial (axial) couplings of fermion $f$ to the $Z$ and $\tilde \Delta^{\rho\mu\nu}_{\rm tot}$ is the triple-gauge vertex after summing over all fermions -- SM and beyond -- in the loop. 
Here we have used $\epsilon_L^\mu(X) \to p^\mu/m_X$ for large $|\vec{p}| \gg m_X$, and so we are implicitly imagining a scenario where
$\sqrt{s}$ of the process is large compared to $m_X$. 
Note that only the transverse part of the $Z$ propagator enters, since $(p+q)_\rho \, \tilde \Delta^{\rho\mu\nu}_{\rm tot} = 0$.

As shown in previous sections, $p_\mu \tilde \Delta^{\rho\mu\nu}_{\rm{tot}}$ 
has the same value whether we consider a St\"uckelberg vector field coupled to global baryon number 
or a gauged baryon number with anomalons much heavier than all of the other physical scales in the process. 
Thus we can evaluate $p_\mu \tilde \Delta^{\rho\mu\nu}_{\rm tot}$ via
\eqnref{wardnonzero} or \eqnref{manomheavy}, yielding
\eqsl{fftoXLstep2}{
i\mathcal M(f \bar f \to X_L \gamma) =  
\frac{i g}{c_W} \frac{\mathcal{A}_B\,e g g_X}{2 \pi^2 c_W} \frac1{m_X} \frac{1}{s - m^2_Z}\bar{v}(k_2)\gamma^{\rho} 
\pr{ q^{\mrm{V}, f}_Z  - q^{\mrm{A},f}_Z \gamma_5 } u(k_1)\, \epsilon^{\rho\nu;p q} \e^*_\nu(q) \,.
}

The details of the calculation of the leading behavior of the squared, polarization-summed 
and initial-state spin-averaged amplitude are given in \appref{fftoXgam}.
We employ \eqnref{ffXgamampschanZ2} with only the AVV terms in the second line
and the massless fermion limit of \eqnref{GsF4lims} to obtain
\eq{
\oln{\abs{\mathcal{M}^2}} \sim 
\frac1{4 (N_c)^2 \pi^4} \pr{ \fracp{g}{c_W}{2} g_X \, e }^2 
\frac{(q^{\mrm{V}, f}_Z)^2 + (q^{\mrm{A},f}_Z)^2}{\pr{s - m_Z^2}^2} 
\cdot \frac{s^2}{4r_X} \pr{1 - \frac{2tu}{s^2}}
\cdot \pr{ \sum_q \k^q Q^q q_Z^{\mrm{A}, q} }^2 \,,
}
where $N_c$ is the number of colors of the initial-state fermions,
$\k^{V,q} = \k^q = 1/3$ is the baryon number of the quarks, 
and $Q^q$ is the electromagnetic charge of the quark $q$.
For up- and down-type quarks, $q_Z^{\mrm{A}, q} = T_3/2$, so $Q^u q_Z^{\mrm{A}, u} = 1/6$ and $Q^d q_Z^{\mrm{A}, d} = 1/12$. 
The entire squared-sum on the right-hand side evaluates to $1/16$ for one generation (including the color factor); 
there is an additional factor of $9$ for three generations. 
This agrees with $\mathcal{A}_B^2 = 9/16$.
The cross section resulting from this amplitude is
\eqsl{fftoXLxsec}{
\sigma(f \bar f \to X_L \gamma) = 
\frac{3}{8\pi} \, \frac1{N^2_c} \, \frac{\alpha^3_{\rm em} \alpha_X}{c^4_W s^4_W} 
\pr{ (q^{\mrm{V}, f}_Z)^2 + (q^{\mrm{A},f}_Z)^2 } 
\frac{(s-m^2_X)^2}{m^2_X\, (s-m^2_Z)^2} \,.
}

This expression already assumes $s \gg m^2_X$ 
(and in the case of gauged baryon number, the masses of any anomalons are much greater than $m_Z$ and $\sqrt{s}$); 
however, there are a couple of further limits that are interesting to explore.
First, consider $s \gg m_Z^2$, with the hierarchy of scales $m^2_X \ll m^2_Z \ll s$. 
In this case, the cross section becomes a constant
\eqsl{fftoXLlimit1}{
\sigma(f \bar f \to X_L \gamma)_{s \gg m^2_Z} = 
\frac{3}{8\pi} \, \frac1{N^2_c} \, \frac{\alpha^3_{\rm em} \alpha_X}{c^4_W s^4_W} 
\pr{ (q^{\mrm{V}, f}_Z)^2 + (q^{\mrm{A},f}_Z)^2 } \frac{1}{m^2_X}.
}
A 2--2 scattering cross section constant in energy implies an amplitude squared that grows as $s$, so an amplitude that grows linearly with energy.

A more interesting limit is $s \ll m^2_Z$, with the hierarchy of scales $m^2_X \ll s \ll m^2_Z$.
In this limit, 
\eqsl{fftoXLlimit2}{
\sigma(f \bar f \to X_L \gamma)_{m_X^2 \ll s \ll m^2_Z} = 
\frac{3}{8\pi} \, \frac1{N^2_c} \, \frac{\alpha^3_{\rm em} \alpha_X}{c^4_W s^4_W} 
\pr{ (q^{\mrm{V}, f}_Z)^2 + (q^{\mrm{A},f}_Z)^2 }  \frac{s^2} {m^4_Z}\frac{1}{m^2_X}.
}
This cross section implies an amplitude squared $\propto s^3$, so an amplitude $\propto s^{3/2}$. To see why this limit is intriguing, let us write the amplitude squared as $\propto \frac{s^2}{m^4_Z}\frac{s}{m^2_X}$. 
If we use the condition $|\mathcal M|^2 = 1$ to set a limit on the cutoff of the theory, we find
\begin{align}
\sqrt{s_{\rm max}} \sim \frac{1}{\alpha_{\rm em}^{1/2} \alpha_X^{1/6}} \Big(\frac{m_X}{m_Z}\Big)^{1/3}\, m_Z \, .
\label{eqn:ffXgammacutoff}
\end{align}
We contrast the above with the result from a four-fermion interaction in the Fermi theory. 
There, the amplitude $\mathcal M(f \bar f \to f \bar f) \sim \frac{s}{m^2_Z}$ (using $m_Z$ instead of $v$ to make the comparison easier and neglecting couplings and numerical factors), implying $\sqrt{s_{\rm max}} \sim m_Z$---a cutoff at the scale of particles we have integrated out. 
Compared to this, the limit from $f \bar f \to X_L \gamma$ is {\em smaller} by a factor of $(m_X/m_Z)^{1/3}$.  
We remark in passing that in \eqnref{ffXgammacutoff}, it is curious
that the cutoff scale of the theory scales as $m_X^{1/3}$ in the same way as the weak gravity
conjecture suggests when $m_Z$ is replaced with $M_{\rm Pl}$ \cite{Craig:2019zkf}.

The situation becomes even more intriguing once we recall that the SM below the weak scale is purely vectorial. 
The triple-gauge vertices formed from loops of fermions with vectorial couplings (VVV in the language introduced in \secref{anomalous}) are zero---stated in the language of gauge anomalies, the theory is anomaly-free. 
As such, $f \bar f \to X \gamma$ cannot exhibit any pathological scaling with respect to $s$ in the limit that we take the weak scale to be infinitely heavy. 
The $1/m^{4}_Z$ scaling on the right-hand side of \eqnref{fftoXLlimit2} satisfies this requirement;
however, unusually, it predicts the scale where perturbative unitarity is violated (using $|\mathcal M|^2 \le 1$) to be parametrically lower than the weak scale. 
If we require that $f \bar f \to X \gamma$ remain valid at least until $m_Z$, this sets a lower limit on the mass of $X$,
\eqsl{mXlimit}{
m_{X, \rm min} \sim \frac{4\,\left[ 3\,\alpha^3_{\rm em}\alpha_X \, \pr{ (q^{\mrm{V}, f}_Z)^2 + (q^{\mrm{A},f}_Z)^2 } \right]^{1/2}}{c^2_W\, s^2_W}\, m_Z.
}
Taking $f$ to be a charged lepton, $m_{X, \rm min} \sim 6\times 10^{-3}\, \sqrt{\alpha_X}\, m_Z$.

The above bound is a function of $\alpha_X$, so it can be made arbitrarily small by sending $\alpha_X \ll 1$; 
in other words, for $\alpha_X \ll 1$, the EFT cutoff implied by \eqnref{ffXgammacutoff} can be pushed above the scale of current experiments. 
It is an interesting and open question as to whether processes such as $f \bar f \to X\gamma$ 
could place bounds on $(\alpha_X, m_X)$ that are competitive with bounds from $Z \to X\gamma$ 
and other electroweak scale processes.

\subsection{$Z\gamma \to Z\gamma$ via $X$ exchange} \label{sec:BIMbaryonnumber}

\begin{figure}[t!]
\begin{center}
\includegraphics[bb = 180 360 440 450]{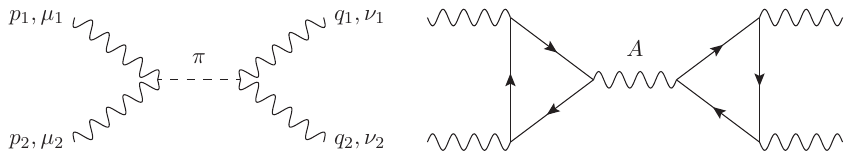}
\end{center}
\caption{$Z\g \to Z\g$ scattering through an $s$-channel $\pi$ (left) or $A$ (right, with 3 cross diagrams not pictured).
}
\label{fig:BIM}
\end{figure}

The final amplitude we calculate using the triple-gauge vertex involves a so-called ``BIM'' process~\cite{Bouchiat:1972iq}, 
the scattering of gauge fields off each other through the exchange of an off-shell St\"uckelberg vector field. 
The original BIM calculation considered the scattering of massless bosons
through a (massive) St\"uckelberg vector field.
For consistency with previous sections, here we specialize the calculation
to the case of a St\"uckelberg vector field coupling to global baryon number,
and so we consider $Z\gamma \to Z\gamma$ through an $s$-channel $X$. 
The diagrams involving $X$ are shown in \figref{BIM}.

There are, of course, additional diagrams from boxes of fermions or $W$ bosons, but these are independent of $g_X$. 
If all loop fermions $\j$ are heavy relative to $\sqrt{s}$, we recover the Euler--Heisenberg Lagrangian from the box diagrams, with $\mathcal M(Z\gamma \to Z\gamma) \sim \frac{s^2}{m^4_\j}$. 
Here we focus on the same scenario considered in the previous subsections,
with only massless SM fermions in the loop.
(We would obtain the same result with gauged baryon number so long as 
the anomalon masses are taken to be much heavier than all other scales). 
Together with the limit $s \gg m_W^2$,
the box diagrams involving $W$ bosons have no bad $s$ behavior%
\footnote{Here we are referring to $(s/M)^n$ behavior at fixed scattering angle, 
where $M$ is some other mass scale in the problem, 
and not to divergences in the limit of forward or backward scattering. 
The latter manifest as ratios of Mandelstam invariants.}
\cite{Jikia:1993tc,Gounaris:1999gh},
so we will neglect them and focus on the contributions from $X$ exchange.

The $X$ exchange occurs through a single diagram stitching together two $Z-\gamma-X$ vertices. 
However, if we write $X^\mu$ as $A^\mu - \partial^\mu \pi/m_X$ and employ gauge fixing as described in \secref{BRST}, there appear to be two diagrams as in \figref{BIM}---one from $\pi$ exchange and one from $A$ exchange, each with gauge dependence.
Feynman rules for these diagrams can be derived from the Lagrangian in \appref{BIM}.

The $A$ exchange piece for $Z_\rho(p_1)\gamma_\nu(q_1) \to Z_{\rho'}(p_2)\gamma_{\nu'}(q_2)$, coming from loops of SM fermions alone, is
\eqsl{BIMzgam1}{
\frac{-i}{s - m^2_X}\tilde \Delta^{\mu\rho\nu}_{\rm SM}\Big( g^{\mu\mu'} -  (1-\xi) \frac{(p_1 + q_1)^\mu (p_2 + q_2)^{\mu'}}{s-\xi\, m^2_X}\Big)\tilde \Delta^{\rho'\mu'\nu'}_{\rm SM}\,,
}
while the $\pi$ piece is
\eqsl{BIMzgam3}{
\frac{i}{s - \xi m^2_X}\Big(\frac{\mathcal{A}_B}{2 \pi^2}\frac{e g g_X}{c_W\, m_X}\Big)^2 \epsilon^{\rho\nu;p_1 q_1}\epsilon^{\rho'\nu';p_2 q_2}.
}
Evaluating the gauge-dependent piece of \eqnref{BIMzgam1} using \eqnref{wardnonzerogaugeB} with appropriate modifications, then its sum with the $\pi$ exchange term,
\eqsl{BIMzgam2}{
& \br{ \frac{i}{s - m^2_X} \frac{(1-\xi) }{s-\xi\, m^2_X} + \frac{i}{m^2_X(s - \xi m^2_X)} }
\Big(\frac{\mathcal{A}_B}{2 \pi^2}\frac{e g g_X}{c_W}\Big)^2 \epsilon^{\rho\nu;p_1 q_1}\epsilon^{\rho'\nu';p_2 q_2}  \\
= &\,  \Big(\frac{\mathcal{A}_B}{2 \pi^2}\frac{e g g_X}{c_W}\Big)^2 \epsilon^{\rho\nu;p_1 q_1}\epsilon^{\rho'\nu';p_2q_2}\frac{i}{s-\xi\, m^2_X}\Big( \frac{(1-\xi)}{s - m^2_X} + \frac 1 {m^2_X}\Big) \\
= &\,  \Big(\frac{\mathcal{A}_B}{2 \pi^2}\frac{e g g_X}{c_W}\Big)^2 \epsilon^{\rho\nu;p_1 q_1}\epsilon^{\rho'\nu';p_2q_2} \frac{i}{m^2_X\,(s - m^2_X)} \,,
}
we see that the $\xi$ dependence cancels. 
Notice that the final result of \eqnref{BIMzgam2} is the same as
just $\pi$ exchange given by \eqnref{BIMzgam3} in the limit
$s \gg m_X^2$ in Landau gauge ($\xi = 0$), 
as required by the longitudinal equivalence theorem in \eqnref{XLpropagatorLandau}.

Assuming $s \gg m_Z^2, m_X^2$, we can neglect the other terms and use \eqnref{BIMzgam2} as an approximation to the full amplitude, deferring a more complete and general calculation to \appref{BIM}. 
Forming a cross section from \eqnref{BIMzgam2} and taking the large-$s$ limit, we find: 
\eqsl{zgamscatter}{
\sigma(Z\gamma \to Z\gamma) \simeq \frac{27}{128\pi^3} \, \frac{\alpha^4_{\rm em}\, \alpha^2_X}{c^4_W\, s^4_W} \frac{s}{m^4_X} + \cdots
}
where the $\cdots$ indicates terms subleading in $s$.

While the diagrams in fig.~\ref{fig:BIM} are reminiscent of longitudinal $W$ scattering in the SM, 
we emphasize that the external $\gamma, Z$ fields in the BIM process are purely transverse. 
In the large-$s$ limit, contracting the vertices above with longitudinal $Z$ polarizations yields zero (for massless SM fermions) via the MCVF.

\section{Discussion} 
\label{sec:discussion}

We have investigated theories with a St\"uckelberg vector field,
emphasizing the systematic approach to constructing an effective
field theory involving $X^\mu$.
We considered several possible interactions of the
St\"uckelberg vector field with the SM or with itself,
identifying the couplings of the longitudinal mode
that lead to scattering amplitudes that grow with energy. 
At tree-level these involve the operators
$(X_\mu X^\mu)^2$, $H^\dagger H X_\mu X^\mu$ and $H^\dagger D_\mu H X^\mu$, 
while the interaction $X_\mu j_{\rm anom}^\mu$ (with
$j_{\rm anom}^\mu$ an anomalous global current) induces
one-loop amplitudes that grow with energy.
The energy growth implies an EFT
with one of these interactions requires a UV cutoff scale
that appears above $m_X$ by an amount that is parametrically 
$1/({\rm coupling})$ of the interaction.
In the specific case of $X^\mu$ coupled to the global baryon current,
we demonstrated that the finite contribution to the fermion triangle diagram
leads to a variety of processes that have longitudinal enhancements
in the small $m_X$ limit, including $Z \to X\gamma$, $f\bar{f} \ra X\gamma$
and $Z\gamma \to Z\gamma$.%
\footnote{The importance of $Z \to X\gamma$ for gauged baryon number
was emphasized in \cite{Dror:2017ehi,Michaels:2020fzj}
along with other FCNC processes involving $K \to \pi X$ and
$B \to K X$ meson decays \cite{Dror:2017ehi}.
Constraints on other $U(1)$s were discussed in \cite{Dror:2020fbh}.}

We performed a detailed analysis of the operator $X_\mu j_{\rm anom}^\mu$.
This interaction is, at first, somewhat puzzling since $X^\mu$ is
not a gauge boson and yet it suggests $X^\mu$ is gauging an anomalous current.
Preskill \cite{Preskill:1990fr} demonstrated that anomalous gauge
theories are simply effective theories with a narrow range
of scales where the EFT is valid.  His analysis emphasized
the UV divergent contributions to the two-point function,
leading to maximum separation between the mass of the gauge
boson of an anomalous theory and the cutoff scale of the theory.
As we have seen, this result holds for theories with
a St\"uckelberg vector field that has no gauge symmetry.
In particular, we demonstrated that the generalized Ward identity
is satisfied if and only if the contributions from both $A^\mu$,
the (fake) gauge boson associated with a (fake) gauge symmetry,
and $\partial^\mu \pi/m_X$ appear in the specific
gauge-invariant combination $A^\mu - \partial^\mu \pi/m_X$. 
Our analysis demonstrates that it is the existence of the
global anomaly, not the gauging of it, that leads to the physical
consequence of scattering amplitudes that grow with energy in the UV\@.
This is reminiscent of \cite{Coleman:1982yg} and may lead to a different
interpretation of anomalies when expressed directly
in terms of on-shell scattering amplitudes.
For example, \cite{Chen:2014eva} recasts the constraints
from anomaly cancellation in terms of on-shell amplitudes
that satisfy unitarity and locality.

As we have seen throughout the paper, the interactions of a
St\"uckelberg vector field that grow with energy can and do arise from
a spontaneously broken $U(1)$ gauge theory with a dark Higgs sector.
In each case, the coefficient of the corresponding operator
depends explicitly on powers of $g_X$, the $U(1)$ gauge coupling.
The cutoff scale of the EFT with the vector boson 
is resolved by dark Higgs exchange, in analogy with the growth
of the scattering of longitudinally-polarized electroweak gauge bosons
in the SM\@.
UV completing an EFT with a St\"uckelberg vector boson using
a dark Higgs sector, in which the vector boson mass
is much smaller than the dark Higgs mass, requires
$g_X \ll \sqrt{\lambda_h}$.
In a $U(1)$ gauge theory in which all $U(1)$ field charges are
order one, this implies that all interactions are suppressed by
powers of the small gauge coupling $g_X$, and in particular,
kinetic mixing arising from integrating out matter that is
charged under the dark $U(1)$ and $U(1)_Y$ is also bounded
by $\epsilon \lesssim g_X e$.  

A St\"uckelberg vector field can be obtained by ungauging a Higgsed theory while holding $m_X$ fixed,
that is, by sending $g_X \to 0$ and thus taking $v_X \to \infty$.
This is distinct from spontaneously-broken gauge theories: 
the limit $m_X \to 0, g_X \to 0$ with the ratio $v_X = m_X/g_X$ held constant does not exist,
demonstrating that a strict interpretation of a theory with a St\"uckelberg vector boson 
does not have anything to do with SSB. There is no Higgs mechanism, no Higgs boson, and so 
the interactions that lead to longitudinally
enhanced scattering amplitudes that grow with energy
have arbitrary coefficients. 
Consequently, a UV cutoff scale of the EFT is inevitable.
Reece \cite{Reece:2018zvv} has suggested that 
weak gravity conjecture arguments \cite{Arkani-Hamed:2006emk}
prevent an arbitrarily small St\"uckelberg mass since  
the limit $m_X \ra 0$ lies at infinite distance in field space.  
It would be interesting to further investigate the constraints
on other parameters of the effective theory of St\"uckelberg
vector bosons using arguments based on embedding the theory
into quantum gravity \cite{Harlow:2022gzl,Draper:2022pvk}.

In the $SU(3)_c \times U(1)_{\rm em}$ effective theory below the 
electroweak scale, all fermion currents are
vectorial with no (gauge or global) anomalies.
Naively, there are no restrictions on coupling an arbitrarily light
St\"uckelberg vector field to any linear combination of these currents.
Of course, the weak interaction explicitly violates some global symmetries,
such as baryon number, so the interactions of $X$ with SM fermion currents
are not purely vectorial. Hence, $X$ will have
scattering amplitudes that grow with powers of $\sqrt{s}/m_X$%
\footnote{An alternative approach in which a vector field interacts only
through higher-dimensional operators was discussed in \cite{Fox:2011qd}.}.
One might think this growth is the same as four-fermion interactions
that also scale with $s/m_W^2$, such that the cutoff scale of the
theory \emph{is} the electroweak breaking scale.
This is not true.  Consider $f\bar{f} \ra X\gamma$ with $X$ coupling to baryon
number.  While there is $s/m_Z^2$ suppression in the amplitude
from $Z$ exchange, 
there is also $\sqrt{s}/m_X$ enhancement from producing
a longitudinally polarized $X$. 
By observing this energy growth in the cross section 
(at energies well below the electroweak scale), 
one could determine whether or not a vector boson 
has longitudinally enhanced couplings.

Finally, we should discuss the status of dark photons that
partly motivated our study of St\"uckelberg vector fields.
In theories where the dark photon Lagrangian arises from a
spontaneously broken $U(1)$ gauge symmetry by a dark Higgs field,
some discussion of the dark Higgs scalar has appeared
(e.g.,~\cite{Ahlers:2008qc,Batell:2009yf,Jaeckel:2010ni,Harigaya:2016rwr,Berger:2016vxi,Battaglieri:2017aum,Darme:2017glc,Mondino:2020lsc}).
Instead, we proclaim that the time is ripe to consider a general set of
interactions that a St\"uckelberg vector field can have
with coefficients that are not dictated by a dark Higgs sector. 
Longitudinally enhanced interactions imply the theory will have a cutoff scale:
within the validity of the effective theory
(i.e., $\sqrt{s}$ less than the cutoff scale as determined by
the longitudinally enhanced scattering processes), 
what phenomenological consequences can arise in the presence 
of these interactions?  
This is an interesting question to explore for more general
vector boson dark matter as well as for dark photon models.

Ultimately our discussion of a St\"uckelberg vector field 
reiterates the lesson of the precarious nature of vector fields in
quantum field theory whose mass is not associated with SSB.
The longitudinal component generically couples to itself or to the SM,
and the presence of these couplings leads to amplitudes that
grow with energy and thus require a cutoff scale for the EFT\@.
There are only two resolutions:
craft the effective theory to have no couplings
of the longitudinal mode, i.e., $X$ coupled only to an anomaly-free
global current, or introduce a Higgs mechanism with a
Higgs boson to restore unitarity of longitudinal vector
boson interactions.  If evidence of a new vector boson
were uncovered in data, we hope our analysis provides a
framework to characterize the effective field theory comprising
the leading interactions of the vector boson independent of its
ultimate UV origin.

\section*{Acknowledgments}

The authors wish to acknowledge that this research was initiated at the Munich Institute for Astro- and Particle Physics (MIAPP) 
conference ``The Weak Scale at a Crossroads: Lessons from the LHC and Beyond" in May--June 2019;
MIAPP is funded by the Deutsche Forschungsgemeinschaft (DFG, German Research Foundation) 
under Germany's Excellence Strategy – EXC-2094 – 390783311. We are grateful to Jordi Eguren and Anton Brekke for identifying mistakes in an earlier version of this paper.
The work of G.D.K. was supported in part by the U.S. Department of Energy under Grant Number DE-SC0011640.
G.L. acknowledges support by the Samsung Science \& Technology Foundation under Project Number SSTF-BA1601-07, 
a Korea University Grant, and the support of the U.S. National Science Foundation through grant PHY1719877.
G.L. is grateful to the University of Toronto for partial support during completion of this work.
The work of A.M. is partially supported by the National Science Foundation under Grant Number PHY-1820860 and PHY-2112540.


\appendix

\section{Form factors in the Rosenberg parameterization of the triangle diagrams}
\label{app:Rosenberg}

In this appendix, we detail the computation of the amplitude of the triple-gauge boson triangle diagrams of \figref{triangle}.
Factoring out couplings, the relevant expressions are of the type in \eqnref{vertexfuncAVV}. 
To compute the finite form factors $F_{3, \ldots, 6}$, we follow the procedure of \cite{Racioppi:2009yxa}.
The denominators on the first and second lines of \eqnref{vertexfuncAVV} can be combined as
\eqs{
& \br{ \pr{(\ell \pm q)^2 - m_\j^2)} \pr{\ell^2 - m_\j^2} \pr{(\ell \mp p)^2 - m_\j^2} }^{-1} \\
= \ & \G(3) \int_0^1 dx \int_0^{1-x} dy \br{\ell^2 \pm 2\ell \cdot k + x q^2 + y p^2 - m_\j^2 + i\vare}^{-3} \,,
}
where $k = x q - y p$;
since we are only interested in the finite form factors, we can make the change of loop momentum $\ell \to \ell \mp k$.
The numerators have terms with up to three powers of $\ell$:
the terms proportional to $\ell^3, \ell^2$ will contribute only to $G^{1,2}$,
and those linear in $\ell$ vanish because they are odd under integration. We use the AVV case as a prototype, finding
\eqsl{VFAVVfinite}{
\left. \G^{\r\mu\nu}_{\rm AVV} \right|\ld{finite} = \int_0^1 & dx \int_0^{1-x} dy \ \G(3) \int_\ell \frac{4i}{(\ell^2 - \D)^3} \\
\Big( & \crl{ (1-x-3y) k^\mu - 2y p^\mu } \e^{\r\nu ; pq} 
+ \crl{ (1-3x-y) k^\nu - 2x q^\nu } \e^{\r\mu ; pq} \\
& - \crl{ (x-y) k^\r + y p^\r + x q^\r } \e^{\mu\nu ; pq} 
\Big) \,,
}
where 
\eql{Delta}{
\D = m_\j^2 - x(1-x) q^2 - y(1-y) p^2 - 2xy \, p \cdot q - i\vare \,.
}
The loop integral evaluates to
\eq{
\int_\ell \frac1{(\ell^2 - \D)^3} = -\frac{i}{(4\pi)^2} \frac1{\G(3)} \D^{-1} \,.
}
To match \eqnref{VFAVVfinite} to the Rosenberg parameterization in \eqnref{RosenbergVF}, we apply the Schouten identity
\eql{Schouten}{
k^\r \e^{\mu\nu\a\b} + k^\mu \e^{\nu\a\b\r} + k^\nu \e^{\a\b\r\mu} + k^\a \e^{\b\r\mu\nu} + k^\b \e^{\r\mu\nu\a} = 0
}
to the last line, which becomes
\eq{
\crl{ (x-y) k^\mu - y p^\mu - x q^\mu } \e^{\r\nu ; pq} 
- \crl{ (x-y) k^\nu - y p^\nu - x q^\nu } \e^{\r\mu ; pq} 
+ \pr{\text{terms in $G^{1,2}_{\rm AVV}$}} \,.
}
The above lead to
\eqsl{RosenbergF}{
F_3 &= - \int_0^1 dx \int_0^{1-x} dy \, y(1-y) \, \D^{-1} \,, \\
F_4 &= - \int_0^1 dx \int_0^{1-x} dy \, xy \, \D^{-1} \,, \\
F_5 &= \int_0^1 dx \int_0^{1-x} dy \, xy \, \D^{-1} \,, \\
F_6 &= \int_0^1 dx \int_0^{1-x} dy \, x(1-x) \, \D^{-1} \,,
}
from which we see that
\eqsl{RosenbergFrelns}{
F_3(p,q) &= -F_6(q,p) \,, \\
F_4(p,q) &= -F_5(p,q) \,.
}

With the $F_{3, \ldots, 6}$ set%
\footnote{Recall, $F_{3, \ldots, 6}$ are independent of the $\mbf{r}_i \in \{\mrm{A},\mrm{V}\}$, so the results of \eqnref{RosenbergF} hold in general and are not specific to the AVV example.}, 
the next step is to express $G^1, G^2$ in terms of $F_{3, \ldots, 6}$ so that the vertex function can be written in terms of the finite form factors.%
\footnote{In the case of massless loop fermions, we note that $F_3$ [$F_6$] suffers infrared divergences if $p^2=0$ [$q^2=0$]. 
This can be seen from \eqnref{Delta} and \eqnref{RosenbergF}.}
To relate $G^1, G^2$ to $F_{3, \ldots, 6}$, we contract $\tilde \Gamma$ with the momenta of 
$A, B, \text{or } C$---respectively, $p_\mu, q_\nu, \text{ or } (p+q)_\rho$. 
From the Rosenberg parametrization of \eqnref{RosenbergVF}, we obtain the following expressions for the momentum-contracted coupling-stripped vertex functions:
\eqsl{RosenbergmomVF}{
(p+q)_\r \, \tilde{\G}^{\r\mu\nu}_{\{\mbf{r}\}}  &= \frac1{\pi^2} \pr{ G^2_{\{\mbf{r}\}}  - G^1_{\{\mbf{r}\}}  } \e^{\mu\nu ; pq} \,, \\
-p_\mu \, \tilde{\G}^{\r\mu\nu}_{\{\mbf{r}\}}  &= \frac1{\pi^2} \pr{ G^2_{\{\mbf{r}\}} - p^2 \, F_3  - p \cdot q \, F_4 } \e^{\r\nu ; pq} \,, \\
-q_\nu \, \tilde{\G}^{\r\mu\nu}_{\{\mbf{r}\}}  &= \frac1{\pi^2} \pr{ G^1_{\{\mbf{r}\}} - p \cdot q \, F_5 - q^2 \, F_6 } \e^{\r\mu ; pq} \,.
}
However, we know that $G^1, G^2$ are not uniquely defined.
To isolate their ambiguities, we first define the triangle vertex function with {\em unshifted} loop momentum 
(i.e., when $a=b=0$ in \figref{triangle})
\eq{
\G^{\r\mu\nu}_{\{\mbf{r}\}} (p, q) \equiv \tilde{\G}^{\r\mu\nu}_{\{\mbf{r}\}} (p, q; z=0, w=0) \,.
}
The difference $\tilde{\G} - \G$ encapsulates the ambiguity from shifting the momentum, 
and for any $\{\mbf{r}\}$ with an odd number of axial couplings, evaluates to \cite{Racioppi:2009yxa}
\eqsl{VFdifshift}{
\br{\tilde{\G} - \G}^{\r\mu\nu}_{\{\mbf{r}\}} 
&= \int_\ell a^\t \pd_{\ell^\t} \mcal{F}^{\r\mu\nu}_{\{\mbf{r}\}} (\ell) 
= \frac{2i\pi^2}{(2\pi)^4} a^\t \lim_{\ell \to \infty} \ell^2 \ell_\t \mcal{F}^{\r\mu\nu}_{\{\mbf{r}\}} (\ell)\, \\
& = \frac1{4\pi^2} \e^{\r\mu\nu\d} a^\d = \frac1{4\pi^2} \e^{\r\mu\nu\d} (z\, p^\d + w\, q^\d) \,,
}
where $\mcal{F}^{\r\mu\nu}_{\{\mbf{r}\}}$ is the integrand in, e.g., \eqnref{vertexfuncAVV} for the AVV case.

We proceed to directly calculate the left-hand sides of \eqnref{RosenbergmomVF}
using the explicit form in \eqnref{vertexfuncAVV}. 
The integrands in each of these contractions can be massaged into terms differing only by a shift in loop momentum, 
which can then be evaluated using the analog of \eqnref{VFdifshift}. For example: 
\eqsl{RosenbergqmomVFexplicit}{
q_\nu \, \Gamma^{\rho\mu\nu}_{\rm AVV} = & \int_\ell \Tr \Bigg\{ \gamma_5\gamma_\rho\frac 1{\slashed{\ell} - \slashed{p} -m_\j} \gamma_\mu \frac 1 {\slashed{\ell} -m_\j} - \gamma_5\gamma_\rho\frac 1{\slashed{\ell} -m_\j} \gamma_\mu \frac 1 {\slashed{\ell} + \slashed{p} -m_\j}\, \\
& \qquad + \gamma_5\gamma_\rho\frac 1{\slashed{\ell} - \slashed{q}-m_\j} \gamma_\mu \frac 1 {\slashed{\ell} + \slashed{p} -m_\j} -  \gamma_5\gamma_\rho\frac 1{\slashed{\ell} - \slashed{p} -m_\j} \gamma_\mu \frac 1 {\slashed{\ell} + \slashed{q} -m_\j} \Bigg\}\,  \\
=  \frac{2i\pi^2}{(2\pi)^4} & \lim_{\ell \to \infty} \Bigg[ p^\t  \ell^2 \ell_\t \Tr\bigg\{ \gamma_5\gamma_\rho\frac 1{\slashed{\ell}} \gamma_\mu \frac 1 {\slashed{\ell} + \slashed{p}} \bigg\} + (p-q)^\t \ell^2 \ell_\t \Tr\bigg\{ \gamma_5\gamma_\rho\frac 1{\slashed{\ell} - \slashed{p}} \gamma_\mu \frac 1 {\slashed{\ell} + \slashed{q}} \bigg\} \Bigg] \, \\
=  \frac 1 {2\pi^2} & \lim_{\ell \to \infty} \frac{1}{\ell^2} \bigg[ p^\t  \ell_\t \epsilon^{\mu\rho;lp} + (p-q)^\t \ell_\t (\epsilon^{\mu\rho;lp} + \epsilon^{\mu\rho;lq} - \epsilon^{\mu\rho;pq}) + \mathcal O(m_\j^2) \bigg] \, \\
= - & \frac{\epsilon^{\rho\mu;pq}}{4\pi^2} \,,
}
where we have used $\ell^{\alpha}\ell^{\beta} \to \ell^2\eta^{\alpha\beta}/4$ to simplify the penultimate line, and the $\epsilon^{\mu\rho;pq}$ term has only one power of $\ell$ in the numerator and therefore vanishes when we take $\ell \to \infty$. 

Below, we list the complete sets of expressions for the AVV case,
\eqsl{RosenbergmomVFAVV}{
(p+q)_\r \, \Gamma^{\rho\mu\nu}_{\rm AVV} &= \frac{\epsilon^{\mu\nu;pq}}{4\pi^2} 4m_\j^2 C_0(m_\j^2) \,, \\
-p_\mu \, \Gamma^{\rho\mu\nu}_{\rm AVV} &= -\frac{\epsilon^{\rho\nu;pq}}{4\pi^2} \,, \\
-q_\nu \, \Gamma^{\rho\mu\nu}_{\rm AVV} &= \frac{\epsilon^{\rho\mu;pq}}{4\pi^2} \,,
}
and the VAV case,
\eqsl{RosenbergmomVFVAV}{
(p+q)_\r \, \Gamma^{\rho\mu\nu}_{\rm VAV} &= 0 \,, \\
-p_\mu \, \Gamma^{\rho\mu\nu}_{\rm VAV} &= -\frac{\epsilon^{\rho\nu;pq}}{4\pi^2} \pr{ 1 + 4m_\j^2 C_0(m_\j^2) } \,, \\
-q_\nu \, \Gamma^{\rho\mu\nu}_{\rm VAV} &= \frac{\epsilon^{\rho\mu;pq}}{4\pi^2} \,.
}
Finally, we can fix $G^{1,2}_{\{\mbf{r}\}}$ by combining \eqnref{VFdifshift} and the above contractions of $(p+q)_\rho, p_\mu, q_\nu$ with unshifted $ \G^{\r\mu\nu}_{\{\mbf{r}\}}$, then equating the sum with \eqnref{RosenbergmomVF}.

\section{Generalized $f\bar{f} \to X\gamma$}
\label{app:fftoXgam}

We examine the amplitude for the $s$-channel (left-hand side) diagram of \figref{ffXgam}:
\eq{
i\mcal{M}_s^{\mu\nu} = \bar{v}(k_2) \crl{ 
\frac{ig}{c_W} \g^\s \pr{ q^{\mrm{V}, f}_Z  - q^{\mrm{A},f}_Z \gamma_5 } \frac{-i}{s-m_Z^2} \pr{\Pi^\infty_Z}_{\s\r}
+ ie Q^f \g^\s \frac{-i}{s} g_{\s\r} } u(k_1) \cdot \tilde{\D}^{\r\mu\nu} \,, 
}
where $Q^f$ is the electromagnetic charge of $f$, 
$(q^{\mrm{V}, f}_Z, q^{\mrm{A},f}_Z)$ are the (vector, axial) charges of $f$ to $Z$,
and the $Z$ propagator in unitary gauge is
\eq{
\pr{\Pi^\infty_Z}_{\s\r} = g_{\s\r} - \frac{(k_1 + k_2)_\s (p+q)_\r}{m_Z^2} \,.
}
The coupling $\j$ must be vector-like. 

Let us isolate the contribution from the intermediate $Z$:
\eq{
\mcal{M}_{s,Z}^{\mu\nu} = -i \frac{g}{c_W} \frac1{s-m_Z^2} \tilde{\D}^{\r\mu\nu} \cdot
\bar{v}(k_2) \crl{ \g_\r \pr{ q^{\mrm{V}, f}_Z  - q^{\mrm{A},f}_Z \gamma_5 } + \frac{2m_f (p+q)_\r}{m_Z^2} \, q^{\mrm{A},f}_Z \g_5} u(k_1) \,.
}
As expected, in the case of the $f\bar{f}Z$ vector coupling, only the transverse part of the $Z$ propagator contributes.
Moreover, we focus on the two cases for which we expect a diverging amplitude:
the AVV and VAV parts of the triangle vertex functions, i.e., 
an axial $Z$ coupling and a vector $X$ coupling or vice versa.
Then
\eq{
\tilde{\D}^{\r\mu\nu} \to \frac{g}{c_W} g_X e \, Q^\j 
\crl{ q_Z^{\mrm{V}, \j} \k^{\mrm{A}, \j} \, \tilde{\G}^{\r\mu\nu}_{\rm VAV} + q_Z^{\mrm{A}, \j} \k^{\mrm{V}, \j} \, \tilde{\G}^{\r\mu\nu}_{\rm AVV} } \,,
}
with $q_Z^\j, \k^\j$ the charge of the fermion $\j$ to $Z, X$ respectively.

Squaring the amplitude, summing over final polarizations, and averaging over initial spins and fermion colors $N_c$, we find%
\footnote{Below, we have assumed that the coupling-stripped vertex functions are real, i.e., that $F_{3, \ldots, 6}$ and $C_0(m_\j^2)$ are real.
For our purposes, we ignore the imaginary parts of these functions, which originate from the possibility of pair production of fermions appearing in the loop.
They can be calculated using the Sokhotski--Plemelj formula applied to the integrands of \eqnref{RosenbergF} and \eqnref{C0} with $\Delta$ from \eqnref{Delta}.
}
\eqsl{ffXgamampschanZ}{
\oln{\abs{\mcal{M}_{s, Z}}^2} = \frac1{4(N_c)^2} & \pr{ \fracp{g}{c_W}{2} g_X \, e }^2 \frac1{\pr{s - m_Z^2}^2 }
\pr{ g_{\mu_1 \mu_2} - \frac{p_{\mu_1} p_{\mu_2}}{m_X^2} } g_{\nu_1 \nu_2} \\
(Q^\j)^2 & \crl{ q_Z^{\mrm{V}, \j} \k^{\mrm{A}, \j} \, \tilde{\G}^{\r\mu_1\nu_1}_{\rm VAV} + q_Z^{\mrm{A}, \j} \k^{\mrm{V}, \j} \, \tilde{\G}^{\r\mu_1\nu_1}_{\rm AVV} } \\
& \crl{ q_Z^{\mrm{V}, \j} \k^{\mrm{A}, \j} \, \tilde{\G}^{\s\mu_2\nu_2}_{\rm VAV} + q_Z^{\mrm{A}, \j} \k^{\mrm{V}, \j} \, \tilde{\G}^{\s\mu_2\nu_2}_{\rm AVV} }
T_{\r\s} \,,
}
where $T_{\r\s}$ is the trace over the external fermion part of the squared-amplitude,
\eqsl{extfermtrace}{
T_{\r\s} = \ & \pr{ (q^{\mrm{V}, f}_Z)^2 + (q^{\mrm{A},f}_Z)^2 } \cdot 4\pr{ \crl{ k_{1\r} k_{2\s} + k_{1\s} k_{2\r} } - \frac{s}2 g_{\r\s} } \\
\ & + \ q^{\mrm{V}, f}_Z q^{\mrm{A}, f}_Z \cdot 8i\e_{\r\s; k_1 k_2} \\
\ & + \pr{q^{\mrm{A}, f}_Z}^2 \cdot 16 r_f \pr{  \frac{s}2 g_{\r\s} + \frac{s}{m_Z^2} \pr{1 + \frac12 \frac{s}{m_Z^2} } (p+q)_\r (p+q)_\s } \,,
}
with $r_i = m_i^2/s$.
We can also simplify the contraction of the external polarization tensors and the triangle vertex functions:
\eqsl{extpoltrans}{
& \pi^4 g_{\mu_1 \mu_2} g_{\nu_1 \nu_2} \G^{\r\mu_1\nu_1}_{\{\mbf{r}\}} \G^{\s\mu_2 \nu_2}_{\{\mbf{r}'\}} \\
= \ & \mcal{F}_T \crl{ \frac1{4r_X} (1-r_X)^2 s g^{\r\s} - \frac1{2r_X} (1-r_X) \crl{ p^\r q^\s + q^\r p^\s } + q^\r q^\s } \\
& - \pr{ 2 r_X G^1_{\{\mbf{r}\}} G^1_{\{\mbf{r}'\}} + (1-r_X) \crl{ G^1_{\{\mbf{r}\}} G^2_{\{\mbf{r}'\}}+ G^2_{\{\mbf{r}\}} G^1_{\{\mbf{r}'\}} } } sg^{\r\s} \\
& + 2 G^1_{\{\mbf{r}\}} G^2_{\{\mbf{r}'\}} \, q^\r p^\s + 2 G^2_{\{\mbf{r}\}} G^1_{\{\mbf{r}'\}} \, p^\r q^\s + 2 G^1_{\{\mbf{r}\}} G^1_{\{\mbf{r}'\}} \, p^\r p^\s + 2 G^2_{\{\mbf{r}\}} G^2_{\{\mbf{r}'\}} \, q^\r q^\s \,,
}
\eqsl{extpollong}{
& -\pi^4 \frac{p_{\mu_1} p_{\mu_2}}{m_X^2} g_{\nu_1 \nu_2} \G^{\r\mu_1\nu_1}_{\{\mbf{r}\}} \G^{\s\mu_2 \nu_2}_{\{\mbf{r}'\}} \\
= \ & \pr{ \mcal{F}_L - G^2_{\{\mbf{r}\}} G^2_{\{\mbf{r}'\}} } \crl{ \frac1{4r_X} (1-r_X)^2 s g^{\r\s} - \frac1{2r_X} (1-r_X) \crl{ p^\r q^\s + q^\r p^\s } + q^\r q^\s } \,,
} 
where $\mcal{F}_{T,L}$ contain products of $F_i$ and $G^{1,2}_{\{\mbf{r}\}, \{\mbf{r}'\}}$, and $\{\mbf{r}\}, \{\mbf{r}'\} \in \{\mrm{AVV}, \mrm{VAV}\}$.

In addition to \eqnref{RosenbergFrelns}, \eqnref{GAVVRosenberg}, and \eqnref{GVAVRosenberg}, we have an additional relation between the form factors in the Rosenberg parameterization by using \eqnref{RosenbergmomVF} and either \eqnref{RosenbergmomVFAVV} or \eqnref{RosenbergmomVFVAV}:
\eq{
r_X \cdot s F_3 = \frac12 + m_\j^2 C_0(m_\j^2) - (1-r_X) s F_4 \,,
}
We can use this relation to simplify the expressions for $G^{1,2}$ in this case:
\eqsl{GAVV}{
G^1_{\rm AVV} &= \frac14 (z+1) - \frac{1-r_X}2 s F_4 \,, \\
G^2_{\rm AVV} &= \frac14 (w+1) - \frac{1-r_X}2 s F_4 + m_\j^2 C_0(m_\j^2) \,,
}
\eqsl{GVAV}{
G^1_{\rm VAV} &= \frac14 (z+1) - \frac{1-r_X}2 s F_4 \,, \\
G^2_{\rm VAV} &= \frac14 (w+1) - \frac{1-r_X}2 s F_4 \,.
}
Then only $F_{4,6}$ and $m_\j^2 C_0(m_\j^2)$ are independent.
In terms of the these functions, the two quantities $\mcal{F}_{T,L}$ in \eqnref{extpoltrans} and \eqnref{extpollong} can then be written as
\begin{align}
\mcal{F}_T = \ & \pr{\frac12 + m_\j^2 C_0(m_\j^2)} \br{\frac12 + m_\j^2 C_0(m_\j^2) - sF_4} 
+ r_X sF_4 \br{\frac12 + m_\j^2 C_0(m_\j^2) - sF_6} \nl
& + r_X^2 sF_4 (sF_4 + sF_6)
+ r_X \pr{ sF_4 - sF_6} \pr{ G^1_{\{\mbf{r}\}} + G^1_{\{\mbf{r}'\}} } \label{eqn:FT} \\
& - \br{\frac12 + m_\j^2 C_0(m_\j^2) - (1-2r_X) sF_4} \pr{ G^2_{\{\mbf{r}\}} + G^2_{\{\mbf{r}'\}} } \,, \nl
\mcal{F}_L = & - \br{ \frac12 + m_\j^2 C_0(m_\j^2) - \pr{ G^2_{\{\mbf{r}\}} + \frac{1-r_X}2 sF_4 } } 
\br{ \frac12 + m_\j^2 C_0(m_\j^2) - \pr{ G^2_{\{\mbf{r}'\}} + \frac{1-r_X}2 sF_4 } } \,. \label{eqn:FL}
\end{align}

We contract \eqnref{extfermtrace} with \eqnref{extpoltrans} and \eqnref{extpollong} and take the limit of massless initial-state fermions, $r_f \to 0$.
Inserting these results back into \eqnref{ffXgamampschanZ}, we obtain the expression at leading-order in $r_X \ll 1$: 
\eqsl{ffXgamampschanZ2}{
\oln{\abs{M_{s, Z}}^2} &\sim \frac1{4N_c^2 \pi^4} \pr{ \fracp{g}{c_W}{2} g_X \, e }^2 \frac1{\pr{s - m_Z^2}^2 } \cdot
\pr{ (q^{\mrm{V}, f}_Z)^2 + (q^{\mrm{A},f}_Z)^2 } \frac{s^2}{4r_X} \pr{1 - \frac{2tu}{s^2}} \\
& \qquad \qquad \cdot (Q^\j)^2 \br{ \sum_{\mbf{r} \in \{\mrm{AVV,VAV}\}} 
q_Z^{\mbf{r}_1, \j} \k^{\mbf{r}_2, \j} 
\pr{2G^2_{\{\mbf{r}\}} - sF_4} }^2 \,.
}
Of the two loop momentum shift parameters, we see from \eqnref{GAVV} and \eqnref{GVAV} that only $w$ appears in the second line.
In order that the Ward identities for the photon and $Z$ boson be satisfied, we must have $w=z=-1$.
Examining the form factor combinations on the second line in the two limits $m_\j^2 \to 0$ and $m_\j^2 \to \infty$, we find:
\eqsl{GsF4lims}{
2G^2_{\rm AVV} - sF_4 &\to \begin{cases} 
-\frac12 & m_\j^2 \to \infty \\ 
-1 & m_\j^2 \to 0 \\
\end{cases} \,, \\
2G^2_{\rm VAV} - sF_4 &\to \begin{cases} 
0 & m_\j^2 \to \infty \\ 
-1 & m_\j^2 \to 0 \\ 
\end{cases} \,.
}

For completeness, we provide expressions for the form factors in the following two limits.
If the loop fermions are infinitely heavy, then at leading order we can discard the $p^2 = m_X^2$ term appearing in \eqnref{Delta}, such that
\eql{heavyloopferm}{
r_\j \to \infty:
\begin{cases}
m_\j^2 C_0(m_\j^2) \to -r_\j \pr{ \mrm{Li}_2\pr{\frac2{1 + i\sqrt{4r_\j - 1}}} + \mrm{Li}_2\pr{\frac2{1 - i\sqrt{4r_\j - 1}}} } \to -\frac12 \\
sF_4 \to \frac12 + m_\j^2 C_0(m_\j^2) \to 0 \\
sF_6 \to 1 - \sqrt{4r_\j - 1} \, \mrm{arccot} \pr{\sqrt{4r_\j - 1}} \to 0
\end{cases}
\,,
}
where $\mrm{Li}_2$ is the dilogarithm function.
If the loop fermions are massless, then $F_6$ suffers an infrared divergence:%
\footnote{As previously mentioned, $sF_6$ also has imaginary part $\pi/4$.}
\eql{masslessloopferm}{
r_\j \to 0:
\begin{cases}
m_\j^2 C_0(m_\j^2) \to 0 \\
sF_4 \to \frac{1 - r_X + r_X\log(r_X)}{2(1-r_X)^2} \\
sF_6 \to \frac14 \pr{1 + 2\log\vare} \qquad (r_X \to 0) 
\end{cases}
\,.
}

\section{Off-shell $X$-exchange amplitudes} 
\label{app:BIM}

In this appendix, we compute the amplitude for $BB \to BB$ scattering (the BIM amplitude after \cite{Bouchiat:1972iq}), for which the diagrams are shown in \figref{BIM}. 
This calculation illustrates the impact of longitudinal enhancement from the triple-gauge vertex when the St\"uckelberg field is off-shell, and it is analogous to $WW$ scattering in the SM.

\begin{figure}[t!]
\begin{center}
\includegraphics[bb = 230 370 400 450]{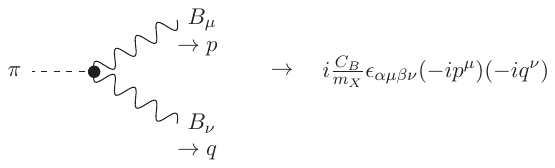}
\end{center}
\caption{Example diagram from dimension-5 counterterms in last line of \eqnref{AB} model.}
\label{fig:PQ}
\end{figure}

A simple setup that accommodates this process is the ``$A$--$B$" model from \cite{Coriano:2008pg}:
this consists of a single Dirac fermion $\j$ with an axial-vector interaction to $A$ and a vector-like interaction to $B$.
The vector field $A$ has a St\"uckelberg-like mass term.
In order to cancel anomalies, the model includes dimension-5 Peccei--Quinn local counterterms coupling the St\"uckelberg scalar field $\pi$ to Chern--Pontryagin densities.
The Lagrangian after performing the $R_\xi$ gauge fixing procedure as in \eqnref{LRxi} is
\eqsl{AB}{
\mcal{L} = -\, & \frac14 B_{\mu\nu} B^{\mu\nu} - \frac14 F_{\mu\nu} F^{\mu\nu} - \frac1{2\xi} \pr{\pd_\mu A^\mu}^2 
+ \frac12 m_X^2 A_\mu A^\mu \\
+ \, & \bar{\j} i \pr{\slashed{\pd} + ie\slashed{B} + ig\slashed{A} \g_5} \j - m_{\j} \bar{\j} \j \\
+ \, & \frac{C_A}{2m_X} \pi F_A^{\mu\nu} \wt{F}_{A\mu\nu} + \frac{C_B}{2m_X} \pi B^{\mu\nu} \wt{B}_{\mu\nu} \,.
}
An example Feynman diagram for the Peccei--Quinn terms in the last line of \eqnref{AB} is displayed in \figref{PQ}.%
\footnote{We can assume that $e \ll g$ such that these diagrams dominate over the standard contribution from one-loop box diagrams of fermions in, e.g., light-by-light scattering.}

The diagram on the left-hand side of \figref{BIM} with $s$-channel $\pi$ exchange evaluates to
\eqs{
i\mcal{M}_1^{\mu_1 \mu_2 \nu_1 \nu_2} &= i\frac{C_B}{m_X} \e^{\mu_1 \a_1 \mu_2 \a_2} (-i p_{1,\a_1}) (-i p_{2,\a_2}) \cdot 
\frac{i}{s - \xi m_X^2} \cdot i\frac{C_B}{m_X} \e^{\nu_1 \b_1 \nu_2 \b_2} (i q_{1,\b_1}) (i q_{2,\b_2}) \\
\mcal{M}_1^{\mu_1 \mu_2 \nu_1 \nu_2} &= - \frac1{s - \xi m_X^2} \frac{C_B^2}{m_X^2} \e^{\mu_1 \mu_2 ; p_1 p_2} \e^{\nu_1 \nu_2 ; q_1 q_2} \,.
}
For the diagram on the right-hand side, we are interested in the part of the amplitude that involves axial couplings of the loop fermions to $A$,
\eqs{
i\mcal{M}_2^{\mu_1 \mu_2 \nu_1 \nu_2} &= \tilde{\D}_{\rm AVV}^{\l \mu_1 \mu_2} (-p_1, -p_2) \cdot 
\frac{-i}{s - m_X^2} \pr{\Pi^\xi_X}_{\l\r} \cdot
\tilde{\D}_{\rm AVV}^{\r \nu_1 \nu_2} (q_1, q_2) \,, \\
\pr{\Pi^\xi_X}_{\l\r} &= g_{\l\r} - (1-\xi) \frac{(p_1 + p_2)_\l (q_1 + q_2)_\r}{s - \xi m_X^2} \,.
}

Rewriting the $R_\xi$ gauge propagator as in the second line of \eqnref{X2pt2},
we can evaluate the $\xi$-dependent longitudinal terms using the MCVF of \eqnref{MCVF}
with $C \to A, A \to B$,
\eq{
\left. \mcal{M}_2^{\mu_1 \mu_2 \nu_1 \nu_2} \right|_\xi = 
\frac1{s - \xi m_X^2} \frac1{m_X^2} \e^{\mu_1 \mu_2 ; p_1 p_2} \e^{\nu_1 \nu_2 ; q_1 q_2}
\cdot \fracp{e^2 g}{4\pi^2}{2} \crl{ (w-z) + 4m_\j^2 C_0 \pr{m_\j^2} }^2 \,.
}

For the gauge dependence to cancel, we must have
\eq{
C_B = \pm \frac{e^2 g}{4\pi^2} (w-z) \,.
}
To satisfy the anomaly-free Ward identities in \eqnref{MCVF} for the $B$ vector bosons, we must choose $w = 1, z = -1$.
The remaining $\xi$-independent amplitude is $\mcal{M}_2$ with the intermediate $A$ propagator in ``unitary" gauge:
\eq{
\mcal{M}^{\mu_1 \mu_2 \nu_1 \nu_2} = \D_{\rm AVV}^{\l \mu_1 \mu_2} (-p_1, -p_2) \cdot 
\frac{-1}{s - m_X^2} \pr{\Pi^\infty_X}_{\l\r} \cdot
\tilde{\D}_{\rm AVV}^{\r \nu_1 \nu_2} (q_1, q_2) \,,
}
which can be broken up into its transverse and longitudinal parts as
\eqsl{BIMtranslong}{
\mcal{M}_T^{\mu_1 \mu_2 \nu_1 \nu_2} &= - \frac{(e^2 g)^2}{s-m_X^2} 
\tilde{\G}^{\l \mu_1 \mu_2} (-p_1, -p_2) \tilde{\G}_{\l}^{\ \nu_1 \nu_2} (q_1, q_2) \,, \\
\mcal{M}_L^{\mu_1 \mu_2 \nu_1 \nu_2} &= - \frac1{s-m_X^2} 
\frac{C_B^2}{m_X^2} \e^{\mu_1 \mu_2 ; p_1 p_2} \e^{\nu_1 \nu_2 ; q_1 q_2} \,.
}
where the subscript AVV is implicit. 
The squared amplitude, averaged and summed over initial and final states, is
\eqs{
\left| \oln{\mcal{M}_T} \right|^2 &= \frac14 \br{ \frac{(e^2 g)^2}{s-m_X^2} }^2 
\tilde{\G}^{\l \mu_1 \mu_2} (-p_1, -p_2) \tilde{\G}^{\r}_{\ \mu_1 \mu_2} (-p_1, -p_2) 
\tilde{\G}_{\l}^{\ \nu_1 \nu_2} (q_1, q_2) \tilde{\G}_{\r \nu_1 \nu_2} (q_1, q_2) \,, \\
\left| \oln{\mcal{M}_L} \right|^2 &= \frac14 \br{ \frac1{s-m_X^2} \frac{C_B^2}{m_X^2} }^2 
\e^{\mu_1 \mu_2 ; p_1 p_2} \e^{\nu_1 \nu_2 ; q_1 q_2} 
\e_{\mu_1 \mu_2 ; p_1 p_2} \e_{\nu_1 \nu_2 ; q_1 q_2} \,, \\
2 \mrm{Re} \pr{ \oln{\mcal{M}^*_L \mcal{M}_T} } &= \frac12 \br{ \frac1{s-m_X^2} }^2 (e^2 g)^2 \frac{C_B^2}{m_X^2}
\e^{\mu_1 \mu_2 ; p_1 p_2} \tilde{\G}^{\l \mu_1 \mu_2} (-p_1, -p_2) 
\e_{\nu_1 \nu_2 ; q_1 q_2} \tilde{\G}_{\l}^{\ \nu_1 \nu_2} (q_1, q_2) \,.
}

Since $p^2 = q^2 = 0$, the evaluation of the vertex functions is simple in the BIM case.
From \eqnref{RosenbergFrelns}, we have both $F_5 = -F_4, F_3 = -F_6$.
Then from \eqnref{GAVV},
\eq{
\left. G^2_{\rm AVV} \right|_{p^2=q^2=0} = - \left. G^1_{\rm AVV} \right|_{p^2=q^2=0} = \frac12 sF_4 \,.
}
Finally, if we have massless loop fermions, then \eqnref{masslessloopferm} implies
\eq{
\left. s F_4 \right|_{m_\j^2 = 0} = \frac{1 - r_X + r_X\log(r_X)}{2(1-r_X)^2} \,. 
}

Then
\begin{align}
\left| \oln{\mcal{M}}_T \right|^2 
&= \frac{(e^2 g)^4}{16} \frac1{\pr{1-r_X}^2} (s F_4)^4
\to \fracp{e^2 g}{4}{4} \,, \nl
\left| \oln{\mcal{M}}_L \right|^2 
&= \frac{C_B^4}{16} \frac1{r_X^2 \pr{1-r_X}^2}
\to \fracp{e^2 g}{4\pi^2}{4} \frac1{r_X^2} \,, \label{eqn:BIMlongitudinal} \\
\mrm{Re} \pr{ 2 \oln{\mcal{M}^*_L \mcal{M}_T} } 
&= \frac{(e^2 g)^2 C_B^2}{4} \frac1{r_X \pr{1-r_X}^2} \pr{ 2G^1 G^2 + \frac{t}{s} (G^1 + G^2)^2 }
\to - \fracp{e^2 g}{2\pi}{4} \frac1{8r_X} \,. \nonumber
\end{align}
where arrows indicate the limit $r_X \to 0$ and we omitted the $AVV$ subscript in the interference term for clarity.

\begin{figure}[t!]
\begin{center}
\includegraphics[bb = 230 370 400 450]{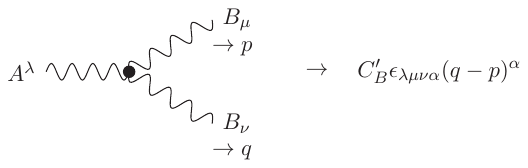}
\end{center}
\caption{Feynman amplitude for Wess--Zumino term in \eqnref{ABWZ}.}
\label{fig:WZ}
\end{figure}

Let us consider adding Wess--Zumino terms to the Lagrangian of \eqnref{AB}.
These are 
\eql{ABWZ}{
\frac12 \e_{\mu\nu\l\r} A^\mu B^\nu \pr{ C'_A F_A^{\l\r} + C'_B F_B^{\l\r} } 
= - \e_{\mu\nu\l\r} A^\mu B^\nu \pr{ C'_A \pd^\r A^\l + C'_B \pd^\r B^\l } \,.
}
For the $BB \to BB$ process, only the $C'_B$ coefficient is relevant;
the amplitude is shown in \figref{WZ}.
Along with \figref{BIM}, we have three additional diagrams where one or both fermion triangle loops in the diagram on the right-hand side of \figref{BIM} is replaced with a three-boson vertex from \figref{WZ}.

We then have
\eqsl{MBIMWZ}{
i\mcal{M}_3^{\mu_1 \mu_2 \nu_1 \nu_2} = C'_B e^2 g \Big\{ &
\e^{\l \mu_1 \mu_2 \a} (p_1 - p_2)_\a \tilde{\G}^{\r \nu_1 \nu_2}(q_1, q_2) 
+ \tilde{\G}^{\l \mu_1 \mu_2}(-p_1, -p_2) \e^{\r \nu_1 \nu_2 \b} (q_1 - q_2)_\b \Big\} \\
\cdot & \frac{-i}{s - m_X^2} \pr{\Pi^\xi_X}_{\l\r} \,, \\
i\mcal{M}_4^{\mu_1 \mu_2 \nu_1 \nu_2} = -(C'_B)^2 & \e^{\l \mu_1 \mu_2 \a} (p_1 - p_2)_\a \e^{\r \nu_1 \nu_2 \b} (q_1 - q_2)_\b \cdot 
\frac{-i}{s - m_X^2} \pr{\Pi^\xi_X}_{\l\r} \,.
}
Again decomposing the $R_\xi$ propagator as in \eqnref{X2pt2}, we find a modified cancellation condition for gauge independence
\eq{
C_B^2 = \pr{ \frac{e^2 g}{4\pi^2} \crl{ (w-z) + 4m_\j^2 C_0 \pr{m_\j^2} } - 2 C'_B }^2 \,.
}
We are left to calculate the squared amplitude that is the sum of eqs.~(\ref{eqn:BIMtranslong})~and~(\ref{eqn:MBIMWZ}), 
the latter with the replacement $\Pi_X^\xi \to \Pi_X^\infty$.
Examining the longitudina pieces as in the second line of \eqnref{BIMtranslong}, we find after using the gauge independence condition above
\eq{
\mcal{M}_L^{\mu_1 \mu_2 \nu_1 \nu_2} = - \frac1{s-m_X^2} 
\frac{C_B^2 - 4(C'_B)^2 + 4C'_B(\pm C_B + 2C'_B)}{m_X^2} \e^{\mu_1 \mu_2 ; p_1 p_2} \e^{\nu_1 \nu_2 ; q_1 q_2} \,,
}
which yields the same result as in \eqnref{BIMlongitudinal} in the relevant limit.

\bibliographystyle{utphys}
\bibliography{stuckeft}

\end{document}